\documentclass{emulateapj}
\usepackage[]{natbib}
\usepackage{verbatim}

\usepackage[colorlinks=true,linkcolor=magenta,citecolor=blue,breaklinks=true]{hyperref}

\journalinfo{Accepted by Astronomical Journal}
\submitted{Accepted by AJ}

\shorttitle{The continuous SFH of NGC~5471}
\shortauthors{Garc\'{\i}a-Benito et al.}

\begin{document}
\title{The 100 Myr Star Formation History of NGC~5471\\ from Cluster and Resolved Stellar Photometry}
\author{Rub\'en Garc\'{\i}a-Benito\altaffilmark{1,2}} 
\author{Enrique P\'erez\altaffilmark{3}}
\author{ \'Angeles I. D\'{\i}az\altaffilmark{2}}
\author{Jes\'us Ma\'{\i}z Apell\'aniz\altaffilmark{3,4}}
\author{Miguel Cervi\~no\altaffilmark{3}}

\altaffiltext{1}{Kavli Institute of Astronomy and Astrophysics, 
Peking University, 100871, Beijing, China; luwen@pku.edu.cn}
\altaffiltext{2}{Departamento de F\'{\i}sica Te\'orica (C-XI), Facultad de Ciencias, 
Universidad Aut\'onoma de Madrid, Ctra. Colmenar Viejo km 15.600, 
E-28049 Madrid, Spain; angeles.diaz@uam.es}
\altaffiltext{3}{Instituto de Astrof\'{\i}sica de Andaluc\'{\i}a, CSIC, Aptdo. Correos 3004, 
E-18080 Granada, Spain; eperez@iaa.es, jmaiz@iaa, mcs@iaa.es}
\altaffiltext{4}{Ram\'on y Cajal Fellow}

\begin{abstract}

We show that star formation in the giant H{\sc ii} region NGC 5471 has been ongoing during the past 100 Myr. 
Using HST/WFPC2 F547M and F675W, ground based JHKs, and GALEX FUV and NUV images, we have conducted a 
photometric study of the star formation history in the massive giant extragalactic H{\sc ii} region NGC 5471 
in M101. We perform a photometric study of the colour-magnitude diagram (CMD) of the resolved stars, 
and an integrated analysis of the main individual star forming clusters and of NGC 5471 as a whole. 
The integrated UV-optical-nIR photometry for the whole region provides two different reference ages, 
8 Myr and 60 Myr, revealing a complex star formation history, clearly confirmed by the CMD resolved 
stellar photometry analysis. The spatial distribution of the stars shows that the star formation in NGC 
5471 has proceeded along the whole region during, at least, the last 100 Myr. The current ionizing 
clusters are enclosed within a large bubble, which is likely to have been produced by the stars that 
formed in a major event $\sim$20 Myr ago.

\end{abstract}

\keywords{H\,{\sc ii} Region, GEHR, star formation history, CMD}

%-----------------------------------------------------------------------------------------------------------------------------------------------------------------
%-----------------------------------------------------------------------------------------------------------------------------------------------------------------
\section{Introduction}

Giant Extragalactic H\,{\sc ii} Regions (GEHRs) are important star formation tracers in the 
chemical evolution of galaxies and can be used as standard candles to estimate 
cosmological distances, thus they have been intensively studied. 
Equally important, they are the best indicators of the conditions 
that lead to massive star formation. These objects are very luminous H$\alpha$ emitters. Their 
H$\alpha$ luminosities, in the range $10^{39}-10^{41}$ erg s$^{-1}$,  imply large 
numbers of ionizing photons, $10^{51}-10^{53}$ $\mathrm{ph}\, \mathrm{s}^{-1}$, an ionizing power 
equivalent to up to several thousands of O5V stars.
Such a massive stellar concentration provides an excellent laboratory to study 
the  modes of massive star formation.

The energy deposited in the interstellar medium in the form of stellar winds, supernova explosions 
and ultraviolet light produce complex stellar and gaseous spatial distributions which change rapidly during 
the few Myr following the onset of the first massive star generation. Although the massive young clusters 
that power the GEHRs are expected to form on a timescale shorter than 10 Myr, some studies in 
regions such as 30 Doradus in the LMC reveal that their stars may not be coeval \citep{wal99}. 
The first population formed in the central regions may have induced star formation in the periphery through 
the action of UV radiation and stellar winds that compress the surrounding molecular clouds. 
Therefore, the study of the interrelation between stars, gas and dust components is highly relevant 
to understand the nature of these regions.

30 Doradus in the Large Magellanic Cloud (LMC) and NGC~604 in M33 are the two largest GEHRs in the Local 
Group and they have been the subject of very detailed analyses. Both are dominated by very young (2-4 Myr) 
ionizing stellar populations, but they present two rather different 
stellar distributions. While NGC~604 is an extended {\it Scaled 
OB Association} (SOBA) with $\sim$200 O+WR massive stars surrounded by nebular 
filamentary structures, 30 Doradus is powered by a Super Star Cluster (SSC) -- a much more compact core 
\citep{maiz01} -- and a halo, with different stellar populations ranging from 2 to 20 Myr \citep{wal97}. 
Both GEHRs include an ongoing star formation with young stars still embedded in their parent molecular cloud. 
The structural similarities and differences between these well-studied and spatially well-resolved Local Group 
GEHRs have made them obvious ideal laboratories in which to study massive star formation, but in order to 
extrapolate our knowledge to distant luminous unresolved starbursts it is critically necessary to 
go a step further and to analyze in detail the somewhat more distant GEHRs just outside the 
neighbourhood of the Local Group.

%--------------------------------------------- FIGURE LARGE SCALE M101
\begin{center}
\begin{figure}
\includegraphics[width=1.0\linewidth]{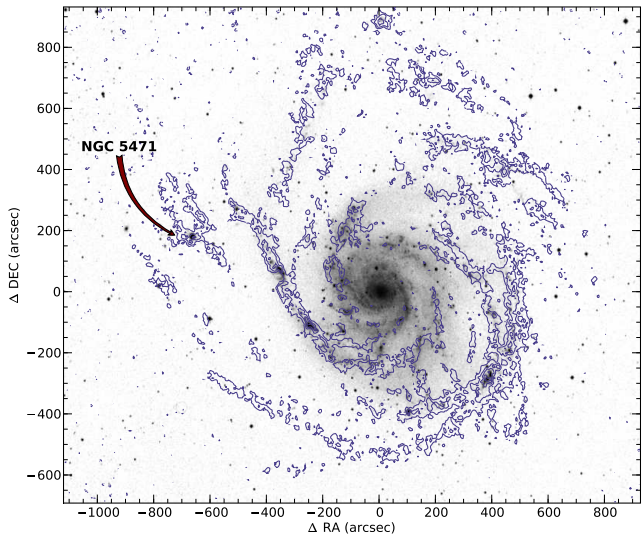}
\caption{ Large scale image of M101. The grayscale is the Digital Sky Survey image and the contour image is 
the H\,{\sc i} from the VLA \citep{braun95}. NGC 5471 is located in the outskirts of the galaxy, in an arm 
concentration of H\,{\sc i}.}
\label{DSS_HI}
\end{figure}
\end{center}
%--------------------------------------------- FIGURE LARGE SCALE M101

M101 (NGC 5457) is a giant spiral galaxy located at a distance of 7.2 Mpc \citep{stetson}
%Stetson, P. B. et al. 1998, ApJ, 508, 491
yielding a linear scale of 34.9 pc/arcsec. This galaxy contains a large number of very luminous 
GEHRs, of which NGC~5471 
($\alpha_{2000}$ = 14$^{h}$04$^{m}$29$^{s}$, $\delta_ {2000}$ = +54\arcdeg23\arcmin48\arcsec; 
l = 101\fdg78, b = +59\fdg63) is one of the outermost,
at a galactocentric radius of about 25 kpc (see figure \ref{DSS_HI}).
The H$\alpha$ morphology of NGC~5471 shows multiple cores with surrounding nebular 
filaments, extending over a diameter of $\sim 17\arcsec$, or $\sim 600$ pc. 
Ground-based photometry shows five bright knots, designated by \citet{skillman}
as components A, B, C, D and E. Component A is as luminous as 
30 Doradus, while components B, C and E are comparable to NGC~604, the two prototypical spatially 
resolved mini-starbursts in the Local Group. Thus NGC~5471 is truly a giant star forming region. 

All these features make of NGC~5471 an excellent candidate for the study of complex star formation, 
being close to the limit distance between the spatially well resolved Local Group 
GEHRs and the more distant unresolved starbursts, thus providing an intermediate 
step in our understanding of very massive star forming regions. 

This paper is organized as follows. 
Section 2 describes the observations and data reduction. 
Section 3 presents the integrated analysis of both the individual large clusters and the whole of NGC 5471,
and the photometry  of the resolved stars. 
The results found in section 3 are elaborated and discussed in Section 4. 
Finally, the conclusions are summarized in Section 5.

%-----------------------------------------------------------------------------------------------------------------------------------------------------------------
%-----------------------------------------------------------------------------------------------------------------------------------------------------------------
\section{Observations and Data Reduction}

We have retrieved ultraviolet NUV and FUV images of M101 from the GALEX archive, as well as broad and 
narrow band optical images for NGC 5471 from the HST archive. Finally, we have obtained near infrared 
JHK images.

%-----------------------------------------------------------------------------------------------------------------------------------------------------------------
\subsection{GALEX data}

The \textit{Galaxy Evolution Explorer} (GALEX)  far-ultraviolet (FUV; $\lambda_{ref}$ = 1530 \AA, $\Delta\lambda$ = 400 \AA) and near-ultraviolet (NUV; $\lambda_{ref}$ = 2310 \AA, $\Delta\lambda$ = 1000 \AA) images of M101 were retrieved  from the 
Nearby Galaxies Survey (NGS).  

The GALEX instrument is described by Martin et al. (2005) and its on-orbit performance by Morrissey et al. (2005). 
GALEX FUV and NUV imaging was obtained with total exposure times of 1041 s in each band. 
The 1$\sigma$ NUV (FUV) sensitivity limit of the GALEX images is 27.6 (27.5) AB mag arcsec$^{-2}$ for these data \citep{bianchi}.
At the distance of M101, the linear scale of 34.9 pc/{\arcsec} implies that the GALEX point spread function (PSF) of 4.6 arcsec 
translates to 160 pc; this is barely sufficient to resolve some of the structure in NGC 5471, and we will only use the GALEX data
to study the integrated photometry.

%-----------------------------------------------------------------------------------------------------------------------------------------------------------------
\subsection{HST/WFPC2 Imaging}

\textit{Hubble Space Telescope} (HST) Wide Field Planetary Camera 2 (WFPC2)
images of NGC~5471 were retrieved from the HST archive. The 
images were taken on 1997 November 1 through two emission-line filters:
F656N (H$\alpha$) and F673N ([S II]), and two continuum filters: F547M 
(Str\"omgren y) and F675W (WFPC2 R) \citep{heyer}. Each camera images onto a 
Loral 800$\times$800 CCD which gives a plate scale of 0\farcs046 pixel$^{-1}$ 
for the PC camera and 0\farcs10 pixel$^{-1}$ for the three WF cameras, with a
readout noise of $\sim$5 e$^{-}$ and a gain of 7 e$^{-}/DN$ for this observations. 
NGC~5471 is centered in the WF3 camera in all images. The scale of the WF 
CCDs at NGC~5471, for which we assume a distance modulus of $(m - M) = 29.3$
\citep{stetson}, is $3.5$ pc pixel$^{-1}$. Table \ref{log} lists the
details concerning the WFPC2 data. 

%--------------------------------------------- FIGURE 4 HST IMAGES
\begin{center}
\begin{figure}%[htbp]
\includegraphics[angle=-90,width=\linewidth]{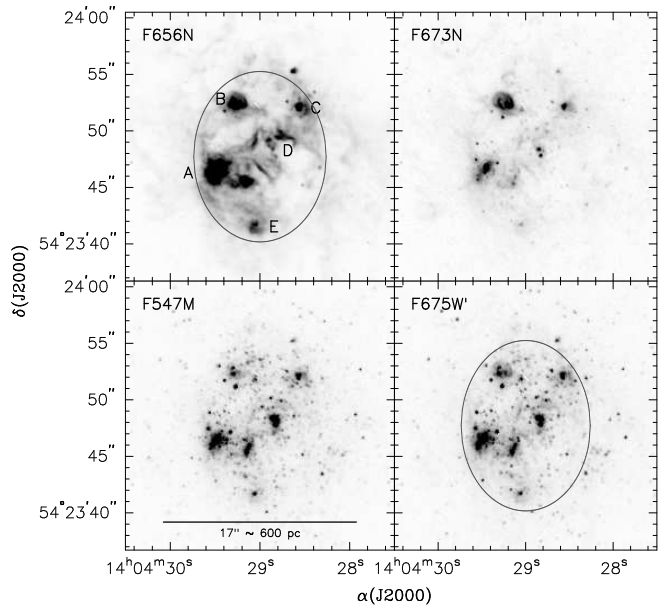}
\caption{
HST WFPC2 images of NGC~5471 in continuum corrected nebular filters  H$\alpha$ (F656N), [SII] (F673N), 
and emission line corrected continuum filters F547M and F675W$^{\prime}$. The five knots from 
\citet{skillman} are marked in the H$\alpha$ image. An ellipse indicating the core of the region is 
drawn in the H$\alpha$ and F675W$^{\prime}$ images.
}
\label{hst}
\end{figure}
\end{center}
%--------------------------------------------- FIGURE 4 HST IMAGES

The HST-pipeline WFPC2 images were subjected to the usual processing 
using the IRAF\footnote{IRAF is distributed by the National Optical 
Observatory, which is operated by the Association of Universities for Research 
in Astronomy, Inc., under cooperative agreement with the National Science 
Foundation.} and STSDAS\footnote{Space Telescope Science Data Analysis System 
(STSDAS) is a product of the Space Telescope Science Institute, operated by 
AURA for NASA} software packages, following the procedures outlined in the HST Data 
Handbook. We used the multiple exposures to correct for the incidence of cosmic rays as well as to  eliminate 
hot pixels and other defects on the CCD. We finally co-added the images to produce deep 
mosaics.

%--------------------------------------------- TABLE 
\begin{deluxetable*}{llcl}
%\tablewidth{0pt}
\tablecaption{Journal of HST/WFPC2 observations of NGC~5471. \label{log}}
\tablehead{
\colhead{Filter} & \colhead{Band} & \colhead{Exposure (s)} & \colhead{Observation ID}
} 
\startdata
F547M & Str\"omgren y & 2 $\times$ 600 &  U4DN030DR, U4DN030ER \\
      &               & 2 $\times$ 100 &  U4DN030FR, U4DN030GR \\
      &               &   20                     &  U4DN030HM \\
F675W & WFPC2 R & 2 $\times$ 400 &  U4DN0308R, U4DN0309R \\
      &               & 2 $\times$ 50   &  U4DN030AR, U4DN030BR \\
      &               & 10                       &  U4DN030CR \\
F656N & H$\alpha$ & 180       &  U4DN0305R \\
      &               & 2 $\times$ 600 &  U4DN0306R, U4DN0307R \\
F673N & [S II] & 3 $\times$ 700 &  U4DN0301R, U4DN0302R, U4DN0303R\\
      &               & 300                     &  U4DN0304R \\
\enddata
\tablecomments{All images were obtained on 1997 November 1 for the cycle 6 
program GO-6928, with You-Hua Chu as PI. The object of study is centered in 
the WF3 camera.}
\end{deluxetable*}
%--------------------------------------------- TABLE 

Given the luminosity of NGC~5471, its H$\alpha$ and [SII] emission contributes 
significantly to the broad band (F675W) image. To eliminate the nebular contribution from the broad band 
and to produce continuum-free H$\alpha$ and [SII] emission line images, we followed the
procedure decribed below. First we multiplied the images by the PHOTFLAM parameter 
obtained by \textit{synphot} to convert from count rates to flux densities, and then divided
them by their respective time exposure. From the SYNPHOT task in STSDAS we determined the filter widths, 
and multiplied the flux densities found by the corresponding filter widths in order to obtain fluxes. 
The correct filter widths are the filter rectangular width RECTW \citep{lurec} defined as
the width in \AA ngstroms of a rectangle with the same total area as the total 
transmission curve and a height given by the peak in the curve. 
We then scaled the H$\alpha$, [S II] and F675W images and performed an iterative 
subtraction to obtain a new, emission line free, broad band image (F675W$^{\prime}$).

Figure \ref{hst} shows the H$\alpha$ image. The five brightest components first noted by 
\citet{skillman} and designated as A, B, C, D and E, are distinctly observed in that figure. 
The presence of a supernova remnant (NGC~5471B) is clearly seen as an intense [SII] emission 
source in the B-component in the F673N image. Previous works have associated 
X-ray sources with supernova remnants (SNRs) in M101. \citet{chen02} 
conclude that the energetic SNR in NGC~5471B \citep[first identified by][]{skillman} was very likely 
produced by a hypernova, an explosion one or two orders of magnitude brighter than the canonical 
10$^{51}$ erg energy explosion associated with a normal SN, although this result has been questioned 
recently \citep{jenkins}.

%-----------------------------------------------------------------------------------------------------------------------------------------------------------------
\subsection{Near-infrared photometry}

JHK$_{\rm s}$ photometry was performed on 2000 April 29, using the 3.58m TNG 
(\textit{Telescopio Nazionale Galileo}) at the Observatorio del Roque de los 
Muchachos in La Palma. The camera used was ARNICA (\textit{ARcetri Near 
Infrared CAmera}), with a CCD detector NICMOS 256$\times$256 HgCdTe and gain 
of 19 e$^{-}$/ADU. The spatial scale was 0\farcs35 
pixel$^{-1}$ and the field of view 90$\times$90 arcsec$^{2}$. The details of these observations 
are given in Table \ref{log2}.

The flux calibration of the images has been done using 2MASS images of NGC~5471, retrieved from 
the 2MASS Extended Source Catalog. We first measured the integrated flux of the 2MASS images 
in a circle containing the region and then, we used that aperture centered in the same coordinates 
and covering the same area in the ARNICA images. We obtained the photometric constants 
comparing the calibrated 2MASS fluxes with those from ARNICA by means of the transformation 
equations of the two systems.

In order to compare the two sets of optical and NIR data, we aligned the HST and TNG images. 
We degraded the WFPC2 images to the spatial scale and seeing of the IR ones, 
then we identified several objects in both data sets and marked them with 
the IRAF \texttt{center} routine. Next, we aligned the images with the 
\texttt{geomap} and \texttt{geotran} routines, using the HST images as 
reference.

%--------------------------------------------- TABLE 
\begin{center}
\begin{deluxetable}{llcl}
\tablewidth{0pt}
\tablecaption{Journal of TNG/ARNICA observations of NGC~5471 \label{log2}}
\tabletypesize{\small}
\tablehead{
\colhead{Band} & \colhead{Time (UT)} & \colhead{Duration (s)} & \colhead{air mass}
}
\startdata
J             & 04:09:00  &   1 $\times$ 60 & 1.38  \\
H            & 01:03:08  &   3 $\times$ 15 & 1.11  \\
K$_{s}$ & 21:59:29  & 10 $\times$   5 & 1.31  \\

\enddata
\end{deluxetable}
\end{center}
%--------------------------------------------- TABLE 

\begin{center}
\begin{figure*}%[htbp]
\includegraphics[width=0.95\linewidth]{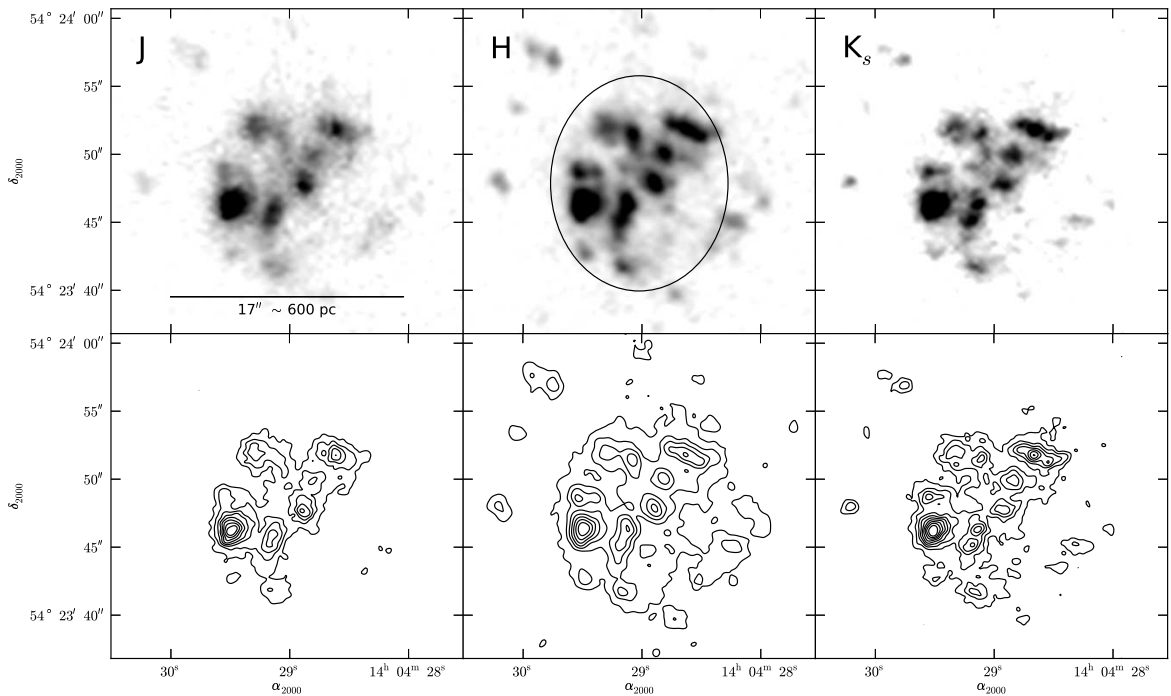}
\caption{
JHK$_{s}$ grayscale images and contour plots. The isophotes range from 20 up to 100 per 
cent of the peak value of the image. An ellipse indicating the core of the region is 
drawn in the H image.
}
\label{arc}
\end{figure*}
\end{center}

Figure \ref{arc} shows the images and contour plots of the infrared bands. The multiple 
knots present in these images are well identified with those in Figure \ref{hst} in the 
F547M and F673W$^{\prime}$ filters.

%--------------------------------------------- FIGURE TNG JHK IMAGES
%--------------------------------------------- FIGURE TNG JHK IMAGES

%-----------------------------------------------------------------------------------------------------------------------------------------------------------
%-----------------------------------------------------------------------------------------------------------------------------------------------------------
\section{Results}

\subsection{Resolved Stellar Photometry}
\label{resolved}

\subsubsection{WFPC2 Photometry}

The stellar photometric analysis was performed with the \textit{HSTphot} 
package \citep{dolpha}. This package is 
specifically designed for use with HST WFPC2 
images. It uses a library of Tiny Tim \citep{krist} undersampled point-spread 
functions (PSFs) for different locations of the star on the camera and of the 
star within the pixel, to center the star and to find its magnitude, 
given in the flight system magnitude. 
It also contains several programs for general reduction procedures. The first 
step is to run the \texttt{mask} task to mask out the bad pixels and other 
image defects identified in the data quality images. The sky computation is 
made by \texttt{getsky}, which takes all pixels in an annulus around each 
pixel, determines the sky value and calculates the sky background map. 
The next step is to run \texttt{crmask} for cosmic ray removal, which uses the 
sky file created by \texttt{getsky}. It has the capability of cleaning 
images that are not perfectly aligned, and it can handle images from different 
filters. To combine multiple images per filter and to produce a final deep 
image per filter, the \texttt{coadd} procedure was used. The final 
step requires the \texttt{hotpixels} procedure, which uses the result from 
\texttt{getsky} and tries to locate and remove all hot pixels. This is an
important step, since hot pixels can create false detections and also,
can throw off the PSF solutions.

The \texttt{hstphot} routine was run on the images in the F547M and F675W$^{\prime}$ bands. This task 
performs stellar PSF photometry on multiple images from different filters, 
including alignment and aperture corrections, as well as PSF modifications 
to correct for errors of geometric distortion via the 
\citet{holt} distorsion correction equations and the 34th row error, noted by 
\citet{sha} (see also \citealt{anderson}), and correction for charge transfer 
inefficiency \citep{dolphb}. There are several parameters to use during 
the run of \texttt{hstphot}. We enabled the determination of the ``local sky'' 
(from now on, \textit{nebular} case), because of the rapidly-varying 
background of NGC~5471, but we also used, in a different run, the usual 
(\textit{stellar}) fit done by \texttt{getsky}. We made a cross-correlation 
between the \textit{nebular} and \textit{stellar} lists of stars in order to match 
them and find the sources detected using both types of sky determination. 
To have a better estimate of the sky for those matched stars, we took as
magnitude the mean of both measurements $m = (m_{nebular} + m_{stellar}) / 2$ 
and as error:

\begin{displaymath}
\sigma_{m} = max\left(\sigma_{m_{n}}, \sigma_{m_{s}}, 
\frac{|m_{n}-m{s}|}{2}\right)
\end{displaymath}

The final list of stars is made up of: i) matched stars; ii) objects from the
\textit{stellar} list not included in the \textit{matched} star list (best detected 
in less nebulous regions); and iii) stars from the \textit{nebular} list with no
companion in the \textit{stellar} list.

We selected ``good stars'' from the \texttt{hstphot} output. Object types were 
classified as good star, possible unresolved binary, bad star, single-pixel 
cosmic ray or hot pixel, and extended object. Less than 10\% of the stars were 
not classified as ``good stars''.  To ensure that we only 
have point-like sources, we also used the ``sharpness'' parameter 
and a measure of the quality of the fit ($\chi^{2}$) to reject false star 
detections in regions with structured nebulosity or artifacts. The final number of stars detected 
in the WF3 chip with these parameters in both filters were 1261.

\subsubsection{Selection Effects}

\paragraph{Completeness}

In order to face the study of the star formation history (SFH) of an object and 
simulate synthetic diagrams, 
it is compulsory to undertake a completeness analysis. 
A thorough description of the completeness analysis can be found in \citet{stetcom} and \citet{apa}. 
We briefly describe below its basis and main steps.

Incompleteness is mainly due to two factors: overlapping of stellar profiles in crowded regions and 
background noise.  These effects are more important for faint stars in dense stellar regions.
Consequently, in a crowded cluster incompleteness will depend on apparent magnitude and 
spatial position. The best way to account for crowding effects is to add artificial stars to the images 
and to subject them to the same sample selection as for real stars. The number 
of injected stars must be sufficiently high as to provide acceptable 
statistical accuracy, but not so high as to change the level of crowding. 
To avoid overcrowding effects, no more than some 10\% of the number of real 
stars should be added as artificial stars. These new artificial images 
(\textit{synthetic frames}) are processed in the same way as the real ones, and 
the magnitudes of all the stars are obtained following the same criteria and 
using the same parameters. At this point, we will have three different tables: \textit{i}) the first one 
containing magnitudes, positions and colors of the real stars (\textit{real 
stars}); \textit{ii}) the second one including all the data for the \textit{injected 
stars}, which are known beforehand; \textit{iii}) the third list containing the data from the 
\textit{synthetic frames}.

The \textit{synthetic table} is then cross correlated with the other two 
(\textit{real} + \textit{injected}), producing a new table 
(\textit{completeness}) containing the stars matched from the artificial list. 
Therefore, in this final list we have the initial and recovered magnitudes 
and colors of the injected stars, as well as their coordinates. The table will also 
contain stars which have not been paired,  and a flag indicating that they have been lost. 
The ratio between the number of simulated stars recovered and injected provides 
the estimate of the completeness factor for each magnitude bin.

\begin{center}
\begin{figure}%[htbp]
\includegraphics[angle=-90,width=\linewidth]{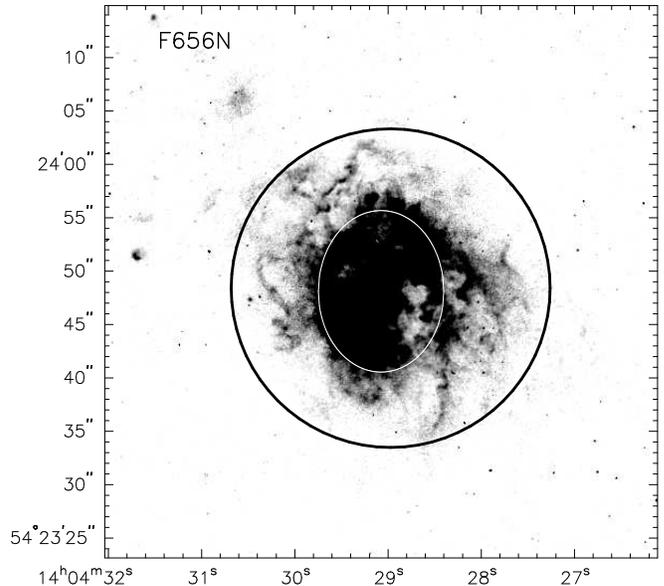}
\caption{Continuum free H$\alpha$ image of NGC 5471. The circular aperture of 15$^{\prime\prime}$ where 
the integrated photometry was performed to obtain the total flux is overplotted together with the 
inner ellipse defining the core of the region.}
\label{apertura}
\end{figure}
\end{center}

A grid of artificial stars was generated on a 2-dimensional CMD and distributed 
according to the flux of the images with an artificial star routine provided 
by \textit{HSTphot}. The parameters of the routine are the minimum and maximum 
of the measured color and magnitude. The magnitude steps used were
multiple of 0.5, while color steps are by 0.25. We asked the routine to 
distribute artificial stars on the CMD in accordance with the number observed. 
Approximately 40,000 fake stars were added in each image (at different trials, in order 
to leave the crowding conditions unaltered) and were given random magnitudes and colors 
in the observed range. We estimated the completeness in two different regimes of crowding: 
inside the core of the region (inner ellipse in Figure \ref{apertura}) and in the rest of 
the image. For the core, the 50\% completeness of the F547M filter luminosity 
function derived on the basis of the CMD data is reached at 24.8 magnitude, while for 
the F675W$^{\prime}$ filter is 24.3 mag. For the rest of the image, the 50\% completeness 
of the F547M filter is reached at 25.2 magnitude, while for the F675W$^{\prime}$ filter 
is 24.7 mag.

%--------------------------------------------------- FIGURE 15arcsec RADIUS APERTURE IN Halpha
%--------------------------------------------------- FIGURE 15arcsec RADIUS APERTURE IN Halpha

\paragraph{Field star subtraction}

Statistical decontamination has been widely used to accomplish the 
field star subtracion of CMDs. The idea is very simple: our region of study,
NGC~5471, consists of the sum of the intrinsic population, IP, and the 
contaminating population, FS, and therefore our CMD shows both populations. 
The field star contribution can be estimated using the CMD of an adjacent 
field FS' containing only foreground and background stars, assuming that it 
is statistically representative of FS. Accordingly, the FS contribution may 
be removed by comparing the local density of stars in the two diagrams 
\citep{bella}:

\begin{displaymath}
CMD(IP+FS)-CMD(FS') \sim CMD(IP)
\end{displaymath}

As NGC~5471 is centered in the WF3, we assume that all of the stars in the 
other chips are field stars. This is not completely true because this GEHR is in the 
outskirts of an M101 arm,  where a few traces of star formation activity can still be found. 
Nevertheless, they are easily detected and no more than 20 stars in the WF2 (the
only chip used to determine the FS population) were found over the chip. The rest of the 
stars showed a homogeneous distribution and were considered a sample of field stars. 
A CMD(IP+FS) was constructed for a region inside a circular apeture of 15$^{\prime\prime}$ 
which includes the core+halo structure of the region.

To perfom the statistical decontamination, we followed the procedure described 
by \citet{mig} and \citet{bella}. For a given star in the CMD(IP+FS) with 
coordinates [$m \pm \sigma_{m}; c \pm \sigma_{c}$],
we count how many stars ($N_{IP+FS}$) can be 
found within the ellipse of axes [$max(\sigma_{m}, 0.1); max(\sigma_{c}, 0.1)$] 
in the original diagram. We also counted the number of stars ($N_{FS'}$) 
found in the field CMD(FS') within the same ellipse centered at the same coordinates. 

The probability $P$ that a given star in the $CMD(IP+FS)$ is actually a 
member of NGC~5471 is defined as follows:

\begin{displaymath}
P = 1 -min \left (\frac{\alpha N_{FS'}}{N_{IP+FS}}, 1.0 \right)
\end{displaymath}

\noindent
where $\alpha$ is the ratio of the area of the IP+FS field (a circular 
aperture of 15$^{\prime\prime}$) to the area of 
the FS' field (in this case, the whole WF2 chip). We can determine the 
probable membership of this star by comparing $P$ with an uniform random 
number  $0 < P' < 1$; if $P' \leq P$ then the star is accepted as a 
member of NGC~5471, otherwise it is rejected and considered as a field star. 
Since this cleaning method is probabilistic in nature, a decontaminating trial  
represents only one out of an infinite number of different realizations of 
the cleaned CMD for NGC~5471. 

%--------------------------------------------- FIGURE OF PROCESS OF CLEANING CMD
\begin{center}
\begin{figure}
\includegraphics[angle=-90,width=\linewidth]{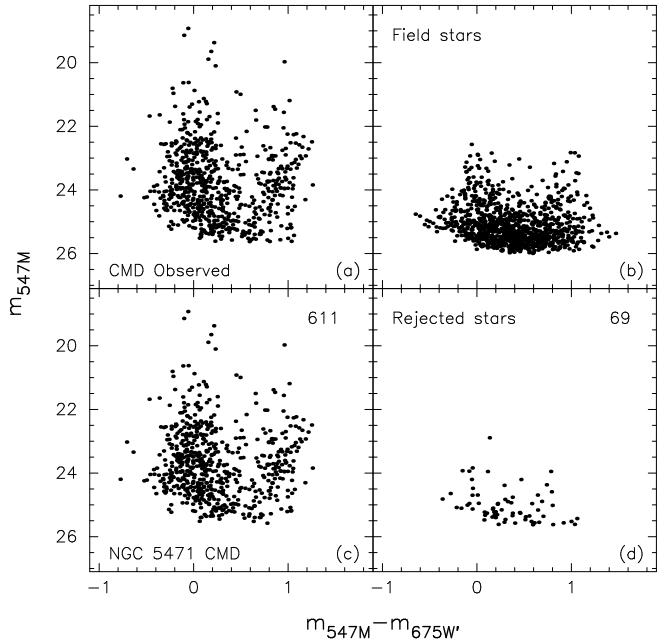}
\caption{
Decontamination of the NGC~5471 observed CMD [panel (a)] using the field stars CMD [panel (b)]. 
The decontaminated NGC~5471 CMD is shown in panel (c) (610 stars) while the CMD of the 
rejected stars is displayed in the panel (d) (73 stars).
}
\label{deconta}
\end{figure}
\end{center}
%--------------------------------------------- FIGURE OF PROCESS OF CLEANING CMD

Figure \ref{deconta} shows the result of this process of decontamination for the CMD of NGC~5471. 
Panel (a) displays the CMD observed (with both intrinsic and field stars), panel (b) the 
field stars, panel (c) the decontaminated CMD of and panel (d) the CMD of 
the stars rejected. Table \ref{stellar} shows the magnitudes of the first 20 stars of the 
resulting clean CMD of panel (c); the full table can be accessed in the electronic edition.

%--------------------------------------------- TABLE 
\begin{deluxetable*}{l c c c c c c c c c c}
\tablecaption{STELLAR PHOTOMETRY \label{stellar}}
\tablehead{
& \colhead{X} &\colhead{Y} & \colhead{R.A.} & \colhead{Decl.} &\colhead{m$_{F547M}$} 
& \colhead{m$_{F675W^{\prime}}$} & \colhead{$m_{F547M} - m_{F675W^{\prime}}$ }\\
\colhead{ID} & \colhead{pixel} & \colhead{pixel} &\colhead{(J2000)} &\colhead{(J2000)} & \colhead{(mag)}
&\colhead{(mag)} & \colhead{(mag)} \\
}
\startdata
1   & 359.56& 562.57& 14:04:27.50& 54:23:41.53& 24.24 $\pm$  0.07& 23.45 $\pm$  0.05&  0.79 $\pm$  0.09\\
2   & 324.75& 512.37& 14:04:27.56& 54:23:47.59& 24.57 $\pm$  0.10& 23.93 $\pm$  0.07&  0.65 $\pm$  0.12\\
3   & 352.44& 544.56& 14:04:27.57& 54:23:43.36& 24.05 $\pm$  0.07& 23.95 $\pm$  0.06&  0.09 $\pm$  0.09\\
4   & 332.62& 514.12& 14:04:27.61& 54:23:46.96& 24.59 $\pm$  0.09& 24.88 $\pm$  0.19& -0.30 $\pm$  0.21\\
5   & 378.56& 569.56& 14:04:27.62& 54:23:39.79& 24.04 $\pm$  0.06& 23.62 $\pm$  0.06&  0.42 $\pm$  0.08\\
6   & 323.33& 496.44& 14:04:27.66& 54:23:48.91& 24.07 $\pm$  0.06& 24.13 $\pm$  0.12& -0.06 $\pm$  0.14\\
7   & 373.56& 553.15& 14:04:27.69& 54:23:41.37& 24.18 $\pm$  0.07& 23.24 $\pm$  0.05&  0.94 $\pm$  0.09\\
8   & 293.29& 455.69& 14:04:27.69& 54:23:53.94& 23.62 $\pm$  0.05& 22.57 $\pm$  0.03&  1.05 $\pm$  0.06\\
9   & 351.67& 523.45& 14:04:27.71& 54:23:45.04& 23.85 $\pm$  0.07& 23.87 $\pm$  0.11& -0.02 $\pm$  0.13\\
10  & 290.75& 449.59& 14:04:27.71& 54:23:54.57& 23.86 $\pm$  0.06& 23.02 $\pm$  0.04&  0.84 $\pm$  0.07\\
11  & 368.80& 540.45& 14:04:27.74& 54:23:42.65& 23.14 $\pm$  0.03& 23.24 $\pm$  0.04& -0.10 $\pm$  0.05\\
12  & 346.68& 513.49& 14:04:27.74& 54:23:46.12& 25.26 $\pm$  0.15& 24.89 $\pm$  0.17&  0.38 $\pm$  0.23\\
13  & 381.63& 555.86& 14:04:27.75& 54:23:40.66& 24.08 $\pm$  0.07& 24.18 $\pm$  0.11& -0.10 $\pm$  0.13\\
14  & 360.90& 530.61& 14:04:27.75& 54:23:43.91& 23.96 $\pm$  0.06& 24.09 $\pm$  0.09& -0.14 $\pm$  0.10\\
15  & 318.82& 469.76& 14:04:27.81& 54:23:51.25& 25.22 $\pm$  0.15& 25.04 $\pm$  0.20&  0.18 $\pm$  0.26\\
16  & 312.47& 461.48& 14:04:27.81& 54:23:52.29& 23.20 $\pm$  0.04& 23.22 $\pm$  0.04& -0.02 $\pm$  0.05\\
17  & 291.51& 435.95& 14:04:27.81& 54:23:55.57& 25.38 $\pm$  0.21& 25.23 $\pm$  0.19&  0.16 $\pm$  0.28\\
18  & 359.87& 517.56& 14:04:27.83& 54:23:44.98& 24.20 $\pm$  0.09& 24.26 $\pm$  0.10& -0.06 $\pm$  0.13\\
19  & 354.84& 509.85& 14:04:27.84& 54:23:45.89& 24.02 $\pm$  0.06& 23.67 $\pm$  0.05&  0.34 $\pm$  0.08\\
20  & 319.56& 460.53& 14:04:27.88& 54:23:51.91& 24.45 $\pm$  0.08& 24.87 $\pm$  0.20& -0.42 $\pm$  0.22\\
\enddata
\tablecomments{Stellar photometry. Table \ref{stellar} is published in its entirety in the electronic edition. 
A portion is shown here for guidance regarding its form and content.}
\end{deluxetable*}
%--------------------------------------------- TABLE 

\subsubsection{Isochrones}
\label{sbsec:iso}

To study the CMD we use the isochrone set that corresponds to the grid by 
\cite{2002AAp...391..195G}\footnote{Available at 
{\tt http://pleiadi.pd.astro.it/}}, covering the age range $6.60<\log t<10.25$ and mass
range 0.15 to 100 M$_\odot$, based on the Z=0.004 (1/5 Z$_\odot$) models (see end of section 
\ref{clphot} for a discussion on the choice of the metallicity) from \cite{2000AApS..141..371G} 
and \cite{1994AApS..106..275B} that include overshooting
and a simple synthetic evolution of TP-AGB \citep{1998MNRAS.300..533G}.
Atmosphere models are from BaSeL 3.1 WLBC 99 corrected library
\citep{2002aap...381..524W} at the same metallicity as the isochrone set.

The procedure is as follows: 
(1) we obtained the magnitudes in the HST filters for all stellar models in the atmosphere grid. 
(2) For each point in the isochrone we obtained the stellar flux in each of the bands by a 
linear interpolation in the $\log g - \log T\mathrm{eff}$ grid. 
(3) The final luminosity was obtained by multiplying by the square radius of the star. 
For $\log g - \log T\mathrm{eff}$ values not covered by the atmosphere grid, 
the boundary values have been used with a caution flag. In terms of the present work, 
these stars do not modify the results obtained.

%--------------------------------------------- FIGURE FINAL CMD WITH ISOCHRONES
\begin{center}
\begin{figure}
\includegraphics[width=\linewidth]{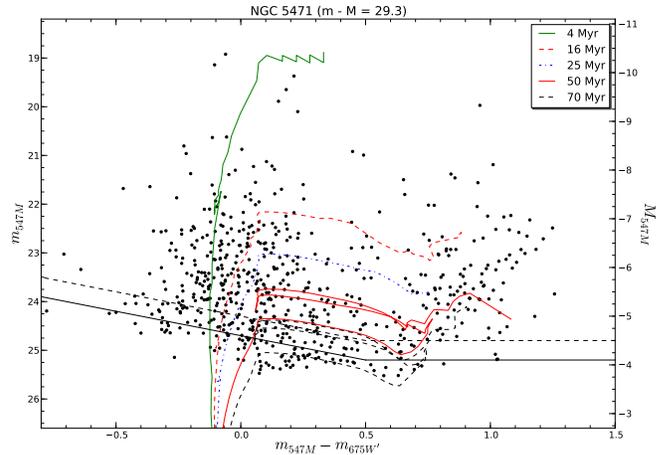}
\caption{
Final CMD corrected for field star contamination, together with five isochrones from 4 to 70 Myr.
The line across the bottom shows the completeness limit at the 50\% level as derived from false 
star tests for the core (dashed line) and for the rest of the image (solid line). 
The right-hand side scale gives the absolute magnitudes corresponding to a distance 
of 7.2 Mpc.}
\label{cmdiso}
\end{figure}
\end{center}
%--------------------------------------------- FIGURE FINAL CMD WITH ISOCHRONES

In Figure \ref{cmdiso} we show the final CMD (corrected for field star contamination).
The 50\% completeness limit is also plotted, and five isochones from 4 to 70 Myr are shown in colour.
The right hand side scale gives the absolute magnitude corresponding to a distance of 7.2 Mpc.

\subsection{Cluster Photometry}
\label{clphot}

\begin{center}
\begin{figure}
\includegraphics[angle=-90,width=\linewidth]{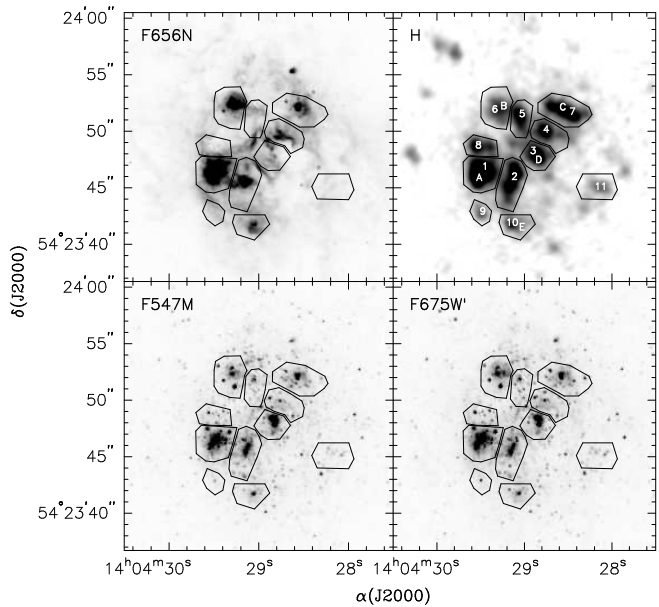}
\caption{
Plot of the eleven polygons, limiting the different areas where the integrated 
magnitudes are computed. The numbers belong to our own definition while the A, B, C, D and E 
letters are the knots identified by \citet{skillman}. The hypernova is located in knot 6.}
\label{poligonos}
\end{figure}
\end{center}

We have derived the integrated magnitudes and colors in 
eleven different apertures covering the main components in the core of NGC~5471. 
The apertures were chosen and defined in the H-band image, maximizing the area in order to include 
the features of all images, since there were some 
small offsets in the position and intensity of the knots in the optical and 
infrared images respectively. Then, the same polygons were used for the J, K$_{\rm s}$ and HST 
images. Polygons were marked by using the IRAF task \texttt{polymark} and the 
integrated magnitudes were computed using the IRAF task \texttt{polyphot}. The 
integrated magnitudes of the HST images are given in the VEGAMAG system. The polygons defined 
are shown in Figure \ref{poligonos}.

%--------------------------------------------- FIGURE DEFINITION OF POLIGONS IN HST AND IN NIR H
%--------------------------------------------- FIGURE DEFINITION OF POLIGONS IN HST AND IN NIR H

The integrated magnitudes and colors derived in each area were corrected for the 
Galactic foreground extinction toward M101 using the color excess 
$E(\bv) = 0.01$ \citep{schle}, although its effect is negligible. The 
absorption coefficients for the 5 bands were derived assuming
$R_{V} = A_{V} / E(\bv) = 3.1$, and using the A$_{\lambda}$ curve from \citet{carde}.
The uncertainty of the individual fluxes have been computed including the Poisson 
photon count error, the uncertainty on the background setting and the photometric 
calibration uncertainty. The results of the integrated photometry are shown in 
Table \ref{magnitude}.

We have also measured the H$\alpha$ flux for each polygon. To measure the integrated 
H$\alpha$ flux of the whole region, we have decontaminated the F656N image from the broad 
band F675W contribution. We have then plotted the contour at 3$\sigma$ from the sky
(with isophotal flux $6.3\times10^{-15}$ erg cm$^{-2}$ s$^{-1}$ arcsec$^{-1}$) and 
found the circular aperture centered in the region that best fits that isophotal 
contour. The radius of this aperture is 15 arcsec from the center (see Figure \ref{apertura}). 
The flux measured inside this circular aperture is 
$(3.65 \pm 0.17) \times10^{-12}$ erg cm$^{-2}$ s$^{-1}$. 
We have used the PHOTFLAM value given by \textit{synphot} and a value of RECTW of 28.3 \AA. 
Measured H$\alpha$ fluxes for each of the polygonal selections and the parameters derived from them 
are given in Table \ref{nebular}. The effective radius (R$_{eff}$) of a polygon is defined as the 
radius of a circular aperture of the same area.

%--------------------------------------------- TABLE 
\begin{deluxetable*}{l c c c c c c}
\tabletypesize{\footnotesize}
\tablewidth{0pt}
\tablecaption{INTEGRATED PHOTOMETRY \label{magnitude}}
\tablehead{
\colhead{Aperture} & \colhead{J} & \colhead{H} & \colhead{K$_{s}$}
& \colhead{m$_{547M}$} & \colhead{m$_{675W}$} & \colhead{m$_{675W^{\prime}}$} \\
}
\startdata
N5471-1    & 16.43 $\pm$ 0.11 & 16.26 $\pm$ 0.12 & 16.03 $\pm$ 0.14 & 16.66 $\pm$ 0.06 & 15.56 $\pm$ 0.06 & 16.30 $\pm$ 0.06 \\
N5471-2    & 16.90 $\pm$ 0.11 & 16.57 $\pm$ 0.12 & 16.48 $\pm$ 0.14 & 17.21 $\pm$ 0.06 & 16.30 $\pm$ 0.06 & 16.94 $\pm$ 0.06 \\
N5471-3    & 17.35 $\pm$ 0.11 & 17.09 $\pm$ 0.12 & 17.12 $\pm$ 0.14 & 17.22 $\pm$ 0.06 & 16.84 $\pm$ 0.06 & 17.11 $\pm$ 0.06 \\
N5471-4    & 17.52 $\pm$ 0.11 & 17.08 $\pm$ 0.12 & 16.99 $\pm$ 0.14 & 18.17 $\pm$ 0.06 & 17.07 $\pm$ 0.06 & 17.77 $\pm$ 0.06 \\
N5471-5    & 17.83 $\pm$ 0.11 & 17.28 $\pm$ 0.12 & 17.27 $\pm$ 0.14 & 18.58 $\pm$ 0.06 & 17.59 $\pm$ 0.06 & 18.16 $\pm$ 0.06 \\
N5471-6    & 17.34 $\pm$ 0.11 & 17.21 $\pm$ 0.12 & 17.00 $\pm$ 0.14 & 17.62 $\pm$ 0.06 & 16.41 $\pm$ 0.06 & 17.12 $\pm$ 0.06 \\
N5471-7    & 16.98 $\pm$ 0.11 & 16.59 $\pm$ 0.12 & 16.47 $\pm$ 0.14 & 17.56 $\pm$ 0.06 & 16.49 $\pm$ 0.06 & 17.10 $\pm$ 0.06 \\
N5471-8    & 17.89 $\pm$ 0.11 & 17.41 $\pm$ 0.12 & 17.39 $\pm$ 0.14 & 18.70 $\pm$ 0.06 & 17.68 $\pm$ 0.06 & 18.22 $\pm$ 0.06 \\
N5471-9    & 19.15 $\pm$ 0.13 & 18.63 $\pm$ 0.13 & 18.63 $\pm$ 0.15 & 20.23 $\pm$ 0.06 & 18.71 $\pm$ 0.06 & 19.54 $\pm$ 0.06 \\
N5471-10  & 18.14 $\pm$ 0.11 & 17.80 $\pm$ 0.12 & 17.57 $\pm$ 0.15 & 18.63 $\pm$ 0.06 & 17.45 $\pm$ 0.06 & 18.22 $\pm$ 0.06 \\
N5471-11  & 18.55 $\pm$ 0.12 & 18.00 $\pm$ 0.12 & 18.11 $\pm$ 0.15 & 19.27 $\pm$ 0.06 & 18.40 $\pm$ 0.06 & 18.84 $\pm$ 0.06 \\
NGC 5471 & 14.14 $\pm$ 0.10 & 13.84 $\pm$ 0.12 & 13.74 $\pm$ 0.14 & 14.67 $\pm$ 0.06 & 13.62 $\pm$ 0.06 & 14.27 $\pm$ 0.06 \\
\enddata
\tablecomments{Integrated magnitudes and H$\alpha$ flux for the eleven regions of NGC 5471.
Magnitudes from HST are given in the VEGAMAG system. Results for
m$_{675W^{\prime}}$ column were measured from the emission line free F675W
image. The last row belongs to the integrated magnitud of the whole region.}
\end{deluxetable*} 
%--------------------------------------------- TABLE 

%--------------------------------------------- TABLE 
\begin{deluxetable*}{l c c c c r}
\tablecaption{NEBULAR PHOTOMETRY \label{nebular}}
\tablehead{
\colhead{Aperture} & \colhead{R$_{eff}$} &\colhead{F(H$\alpha$)} & \colhead{L(H$\alpha$)} & \colhead{Q(H$^{0}$)} &
\colhead{M\,{H\sc ii}} \\%& \colhead{EW(H$\alpha$)} \\
& \colhead{pc} &\colhead{$10^{-13}$ erg s$^{-1}$ cm$^{-2}$} &\colhead{$10^{39}$ erg s$^{-1}$} 
& \colhead{$10^{51}$ s$^{-1}$} &\colhead{$10^{4}$ M$_{\odot}$} \\%& \colhead{A} \\
}
\startdata
NGC5471-1   &    61.1 &   6.63 $\pm$   0.31 &    4.1 &    4.1 &   13.0 \\
NGC5471-2   &    60.9 &   3.07 $\pm$   0.14 &    1.9 &    1.9 &    6.1 \\
NGC5471-3   &    47.0 &   0.93 $\pm$   0.04 &    0.6 &    0.6 &    1.9 \\
NGC5471-4   &    51.2 &   1.59 $\pm$   0.07 &    1.0 &    1.0 &    3.2 \\
NGC5471-5   &    46.0 &   0.79 $\pm$   0.04 &    0.5 &    0.5 &    1.6 \\
NGC5471-6   &    58.9 &   2.42 $\pm$   0.11 &    1.5 &    1.5 &    4.8 \\
NGC5471-7   &    63.1 &   2.34 $\pm$   0.11 &    1.5 &    1.4 &    4.7 \\
NGC5471-8   &    43.5 &   0.72 $\pm$   0.03 &    0.5 &    0.4 &    1.4 \\
NGC5471-9   &    34.0 &   0.38 $\pm$   0.02 &    0.2 &    0.2 &    0.8 \\
NGC5471-10 &    46.2 &   1.22 $\pm$   0.06 &    0.8 &    0.8 &    2.4 \\
NGC5471-11 &    53.6 &   0.31 $\pm$   0.01 &    0.2 &    0.2 &    0.6 \\
NGC 5471     &    525  &   36.5 $\pm$   1.70 &     23 &     22 &     73 \\
\enddata
\tablecomments{Integrated H$\alpha$ flux and derived parameters for the eleven regions of NGC 5471. R$_{eff}$ 
is the effective radius of the polygon. The last row belongs to the integrated values of the whole region.}
\end{deluxetable*}
%--------------------------------------------- TABLE 

Since GALEX images do not have the necessary resolution, we were not able to get the fluxes for 
each polygon. As in the case of the nebular flux, we measured the FUV and NUV fluxes inside 
15\arcsec\ radii apertures centered in NGC 5471. We obtain a FUV flux of 
$(9.9 \pm 0.1) \times 10^{-14}$ erg cm$^{-2}$ s$^{-1}$ \AA$^{-1}$ ($14.19 \pm 0.05$ AB magnitudes) 
and a NUV flux of
$(4.5 \pm 0.1) \times 10^{-14}$ erg cm$^{-2}$ s$^{-1}$ \AA$^{-1}$ ($14.24 \pm 0.05$ AB magnitudes). 

In order to use these values in the analysis, we need to transform AB magnitudes to the VEGAMAG system. 
The transformation formula is \citep{maizp}:

\begin{equation}
AB_{p} = (m_{p} - ZP_{p}) + AB_{p}(Vega)
\end{equation}

\noindent where AB$_{p}$ is the magnitude of the object in the AB system and filter passband $p$,  m$_{p}$ the 
magnitude of the object in filter $p$ using Vega as reference spectral energy distribution (SED), ZP$_{p}$ is the zero 
point of the $p$ filter in the Vega system and AB$_{p}$(Vega) is the AB magnitude of Vega in filter $p$. 
This magnitude is given by:

\begin{equation}
AB_{p}(Vega) = -2.5\,log_{10} \left( \frac{\int P_{p}(\lambda)f_{\lambda,Vega}(\lambda)\lambda d\lambda}
{\int P_{p}(\lambda)(cf_{\nu,AB}/\lambda)d\lambda} \right)
\label{abv}
\end{equation}

\noindent where P$_{p}$($\lambda$) is the dimensionless filter sensitivity function of filter $p$, f$_{\lambda,Vega}(\lambda)$ is 
the Vega spectrum and f$_{\nu,AB}$ = 3.63079 $\times$ 10$^{-20}$ erg s$^{-1}$ cm$^{-2}$ Hz$^{-1}$ is the reference 
spectral energy distribution of the AB system.

We have used FUV and NUV filter passbands to calculate the AB magnitudes of Vega by means of equation \ref{abv}. As the 
zero points for GALEX are ZP = 0.000 in both bands, we have

\begin{equation} 
m_{FUV} = AB_{FUV} - 2.073
\end{equation}
\begin{equation} 
m_{NUV} = AB_{NUV} - 1.647
\end{equation}

Finally, the GALEX VEGAMAG magnitudes of NGC~5471 used below are m$_{FUV}$ = 12.12 
and m$_{NUV}$ = 12.59.

Although a complete analysis of the stellar populations is better characterized by the 
CMD study, integrated photometry provides additional information from wavelengths where 
CMD studies are not feasible.
However, the usual methodology to infer properties from clusters based on their integrated light 
is not trivial for low mass clusters and we need to be sure that the precedure is valid before 
the use of any analysis tool. The first trivial test is that cluster luminosity fulfills the 
bolometric assumptions implicit in the synthetic photometry \citep[cluster luminosities larger 
than the {\it Lowest Luminosity Limit} (LLL),][]{CL04}. This basically establishes when the 
cluster luminosity in a given wavelength/filter {\it may be} produced by just an individual 
star\footnote{The LLL just establishes this possibility, but it does not provide more information 
about whether it is really the case.}. The mean luminosity of a cluster obtained by population 
synthesis models can be safely used when the cluster is brighter than 10 times the LLL, or 2.5 in 
magnitude units \citep{CL04}.

%--------------------------------------------- FIGURE LLL
\begin{center}
\begin{figure}
\includegraphics[width=1.0\linewidth]{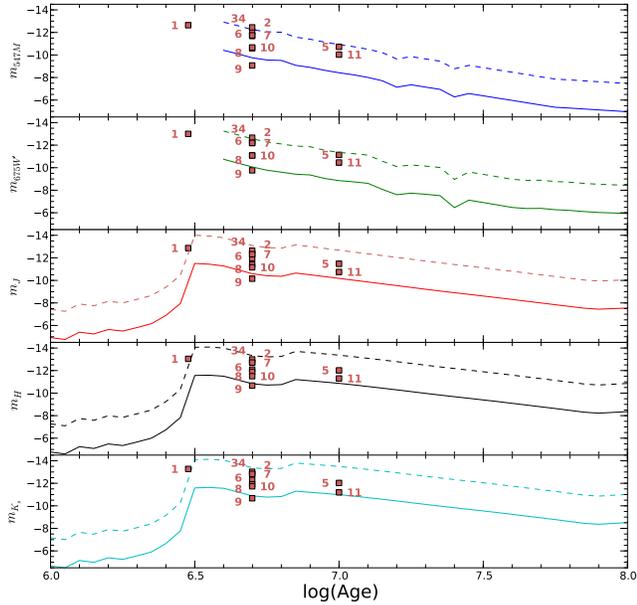}
\caption{Lowest-luminosity limit for the optical and infrared filters used in the cluster photometry.
Each cluster is located at the age resulting from the CHORIZOS analysis. The continuous line indicates 
de LLL and the dotted line the ``safety'' limit of LLL+2.5. 
}
\label{lll}
\end{figure}
\end{center}
%--------------------------------------------- FIGURE LLL

The results of the LLL analysis are shown in figure \ref{lll}, where the ages are just first order 
estimations when LLL and additional limitations in the analysis are not taken into account. 
The figure shows that, with the exception of a few clusters, most of the clusters do not fulfill 
the 10$\times$LLL safety limit. Thus, formally, the analysis based on the mean values obtained from 
synthesis models should be considered as preliminary and used with a cautionary note, as some could 
be biased by contributions from individual stars.

With this limitation in mind and taking into account that the results could be reinterpreted 
we have analyzed the integrated photometric measurements %end{MCS}
with the help of CHORIZOS\footnote{
The latest version can be obtained from \texttt{http://www.iaa.es/$\sim$jmaiz}}, 
a code written in IDL by \citet{maiz}. This code compares photometric data with model spectral 
energy distributions (SEDs), calculating the likelihood for the full parameter range specified, 
allowing for the identification of multiple solutions. 
CHORIZOS can handle SED families for stellar atmospheres and the mean SED obtained by synthesis 
models\footnote{Synthesis models provide an average SED with a relative variance equal 
to zero only in the case of clusters with an infinite number of stars \cite{CL06}.}
\citep[Starburst99;][]{sb99} and uses $\chi^2$ minimization 
to find all models compatible with the observed data in a (2 intrinsic +2 extrinsic)-dimensional 
parameter space.  It includes a wide list of filters for different systems 
and allows for published magnitudes to be used
directly. For a set of objects with $M + 1$ measured magnitudes 
(with independent uncertainties),  
derived $M$ independent colors, 
and a set of N parameters with $M > N$, no strict solutions, but only approximate ones, 
can be found since the problem has more equations or sets of model parameters that 
produce a given color than model parameters. Hence, the code finds
the N-volume of approximate solutions by using $\chi^2$ minimization by the comparison 
with synthetic photometry. In the case of stellar clusters
the synthetic photometry is obtained from the average SED of population synthesis models. 
In these cases, and in addition to the success of the 10$\times$LLL  test,
the $\chi^2$ minimization procedure is valid as far as (a) the synthetic photometry used 
has an intrinsic variance due to size of sample effects much lower than the uncertainty in 
the data \cite[][]{CL06}, (b) size of sample effects do not produce biased synthetic 
photometry \cite[][]{CVG03}, and (c) the average values obtained from the models and used in 
the fit are representative of the intrinsic underlying distribution, that is, the underlying 
distribution of integrated luminosity have a gaussian-like shape and the intrinsic variance can 
be translated to gaussian-like confidence intervals and  $\chi^2$ minimization procedures are 
valid \cite[][]{CL06}\footnote{This requirement is actually linked with the 10$\times$LLL 
and the unbiased photometry requirement.}; this last requirement implies the analysis of 
high-order moments of the distribution and a first order estimation of the mass and 
age of the cluster in absence of external constrains and once the 10$\times$LLL requirement 
is fulfilled by the observational data.
Although, from the LLL test, we know that the results from CHORIZOS should be interpreted with 
caution for the individual clusters analyzed (but valid for the cluster as a whole), we have 
used it to obtain a first order estimation of the effects of extinction and reddening which would 
modify the LLL test.

The first step of the program is to read the unreddened SED models, extinguish them according 
to a given extinction law, and obtain the synthetic photometry.
To run the program, CHORIZOS requires a file with the photometry data and a series of 
input parameters, such as the model family, understood as the collection of SED from stellar 
atmosphere models or synthetic stellar populations, and the range of the parameters to be used.
Then, CHORIZOS reads the model photometry from the tables calculated previously, interpolates, 
calculates the likelihood for each grid point, and writes the results for each of the four 
parameters.

In the case of our study, we have used the STARBURST99 models \citep{sb99} with the implementation 
by \citet{smith02} which includes a grid of blanketed stellar atmospheres at different metallicities.
We fixed two parameters: the known metallicity of NGC~5471 and the type of dust, leaving 
unconstrained the amount of extinction and the age. 

Regarding extinction, CHORIZOS works with $E(4405-5495)$ and 
$R_{5495}$, which are the monochromatic equivalent to $E(B-V)$ and $R_{V}$ respectively, the 
values 4405 and 5495 referring to the central wavelength (in \AA) of the $B$ and $V$ filters. 
We used the standard type of dust characterized by a value $R_{5495} = 3.1$.
The metallicity of the region was assumed to be traced by the oxygen abundance. Recent 
derivations from detailed spectrophotometric studies give a mean value of 12+log(O/H) of 
8.06 $\pm$ 0.03 for components A, C and D; a slightly lower abundance of 7.94 $\pm$ 0.03 is 
found for component B \citep{kbg03}. These values correspond to about 0.25 the solar oxygen 
abundance, if the value by \citet{asplund05} for the solar oxygen abundance is assumed. 
Therefore we used the value $Z = 0.2 Z_{\odot}$ in CHORIZOS.

%-----------------------------------------------------------------------------------------------------------------------------------------------------------------
%-----------------------------------------------------------------------------------------------------------------------------------------------------------------
\section{Discussion}

In this section we discuss the results of the photometry in terms of the resolved stellar 
photometry in the CMD and in terms of the integrated properties, both of the individual 
clusters defined in section \ref{clphot} and of the whole star forming complex. We attempt to 
bring all the results into a coherent picture of the past and present star formation processes 
in NGC 5471.

%-----------------------------------------------------------------------------------------------------------------------------------------------------------------
\subsection{Cluster Analysis}

Analysing the integrated properties of a complex region like NGC 5471 is an exercise that %one 
has to be undertaken with care, because the integrated photometry includes a wide range of regions 
with different cluster masses, ages and extinction values. Thus the result of synthesizing 
the integrated ultraviolet-to-NIR photometry with CHORIZOS, which finds the most likely 
single stellar population (SSP) under 
unconstrained inputs for mass, age and extinction,
is only of general interest as a guiding signpost. If we 
use only the optical-to-NIR photometry, CHORIZOS finds a most likely SSP with relatively 
high extinction and a young age 3 $\pm$ 2 Myr. If we include the FUV and NUV GALEX photometry 
then CHORIZOS finds two loci of maximum likelihood for a single SSP, one that corresponds 
to an age of 8 $\pm$ 2 Myr with moderate extinction and a second maximum at an age 
60 $\pm$ 15 Myr and very low extinction. This result suggests that a significant fraction of the 
GALEX flux integrated within the 15 arcsec radius aperture from NGC 5471 is produced by a post-nebular 
stellar population of B stars \citep[see also][]{waller97}.

Three previous studies have modelled the age of NGC 5471 using ultraviolet-to-optical spectra.
Rosa \& Benvenuti (1994) use an HST/FOS spectrum through an aperture of 4.3 arcsec around 
knot A. Mas-Hesse \& Kunth (1999) use an IUE spectrum through an aperture of 10$\times$20 
arcsec and the FOS optical spectrum of Rosa \& Benvenuti (1994). Pellerin (2006) uses a FUSE 
spectrum through an aperture of 4$\times$20 arcsec. All three works find that the UV spectrum 
of NGC 5471 can be explained with a young burst of less than about 5 Myr and a relatively low 
extinction. 
These results are consistent with the detection of Wolf-Rayet (WR) stars in NGC 5471 \citep{mashesse99}. 
However these previous works may not be sensitive to an older population since they are based on the 
analysis of UV continuum and absorption lines (Si {\sc{iv}} and {C \sc{iv}}) and optical emission 
lines as primary age indicators, and are thus biased towards younger knots with brightest 
ultaviolet fluxes;
while the integrated GALEX photometry
through a large aperture is more sensitive to the 
extended, fainter, more evolved stellar population\footnote{Recall that the 1$\sigma$ 
detection for the GALEX data is AB$_{\rm FUV}$=27.6 mag arcsec$^{-2}$}.
Indeed, \cite{Meurer95}, \cite{Maoz96}, and \cite{Chandar05} have found that the relative 
contribution from the cluster stellar population to the total UV flux in starbursts is only a fraction 
in the range $\sim$20\% to $\sim$40\%, the majority of the UV flux coming from a diffuse fainter stellar 
population that is most prominent when the total flux is integrated, just as in the case of NGC 5471. 
In addition, the analysis performed with CHORIZOS focus the results on the continuum shape.
This discrepancy between ages inferred from emission lines and from the continuum points towards a 
complex star formation history in NGC 5471 as we show below.

The 
integrated
photometry analysis of the clusters identified in section \ref{clphot} 
may be affected by size of sample effects due to the failure of the 10$\times$LLL test in some of 
the clusters. However, we present here the analysis to show, as a case of study, the possible 
interpretations. The values quoted must not be taken, in any case, as final values for cluster 
ages.
%--------------------------------------------- FIGURE Halpha vs. J-H vs. J-V FLUX COLOR COLOUR DIAGRAM
\begin{center}
\begin{figure}
\includegraphics[width=1.0\linewidth]{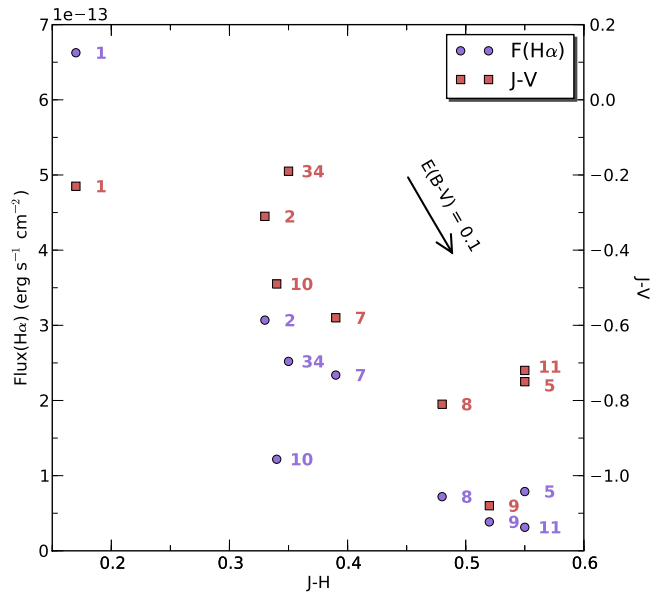}
\caption{H$\alpha$ and J-V versus J-H plot for the clusters identified in section \ref{clphot} 
(V stands for the 547M filter). A clear correlation of decreasing H$\alpha$ with redder 
colors points to an aging effect within the ongoing starburst phase.
}
\label{Ha_JH_JV}
\end{figure}
\end{center}
%--------------------------------------------- FIGURE Halpha vs. J-H vs. J-V FLUX COLOR COLOUR DIAGRAM

Figure \ref{Ha_JH_JV} shows a color-color-H$\alpha$ plot of the individual clusters. 
The hypernova knot is not shown, and knots 3 and 4 have been combined into a single 
knot 34\footnote{The nature of knots 3+4 is not clear, it may be that both are part of the 
same cluster that has evolved sufficiently for the stellar winds to evacuate its center from 
ionized gas. Further higher spatial resolution IR and UV imaging is needed to establish 
more details.}. There is a clear correlation in the sense that redder knots have less H$\alpha$ flux.
The flux of H$\alpha$ can decrease due to the aging of the burst or to different clusters having 
different masses, less massive bursts producing fewer ionizing photons\footnote{Extinction may 
also play a small role but values significantly larger than reported in the literature, 
which are lower than E(B-V)=0.14, would be required.}. 
This latter could be the case, including a relationship with color, when strong size of sample 
effects are present and just a few stars are responsible for most of both the H$\alpha$ 
and continuum emission (over-luminous case) or by the absence of the most massive stars 
which mimics variations in the upper mass limit of the Initial Mass Function 
\citep[under-luminous case; e.g.][]{VCL10}.

Knot 1 is the brightest in H$\alpha$ and the age estimations obtained by CHORIZOS 
give the youngest age, 3 $\pm$ 2 Myr, with the lowest extinction. For knots 2, 3, 4, 7, 8, 9, 10 
CHORIZOS estimates an intermediate age, 5 $\pm$ 2 Myr, with an intermediate extinction, except 
may be knot 9 which has some indications for a somewhat higher extinction. Knots 5 and 11 
have older age estimations, 9 $\pm$ 3 Myr. However, except for Knot 1, 2 and 34, the age estimations 
can also be explained by size-of-sample effects with under-luminous case. In these cases, age 
estimations, although apparently consistent, should not be considered as final, 
pending a more complete analysis with additional data sets.  

In summary, from the cluster analysis of individual bright knots and from the integrated 
photometry of the whole complex there are clear indications that there is a complex star formation 
in NGC 5471 event, with indications of an older population of B stars and a 
luminous young ionizing cluster. This will be well established with the analysis of the CMD 
in the next section.

%-----------------------------------------------------------------------------------------------------------------------------------------------------------------
\subsubsection{Mass of the ionized gas}
\label{sec:massion}

The H$^+$ mass  and the mass of the ionizing cluster may be derived from the H$\alpha$ flux. 
The published values of internal extinction for NGC~5471 differ from different authors,
but they are lower than about E(B-V)=0.14 on average \citep{skillman,tpeim,keni96,luri,esteban}. 
In NGC 604 \cite{Maiz04} have estimated that only about 10\% of the H$\alpha$ flux is missing in 
the higher opacity zones. Therefore, we have not attempted to correct the measured H$\alpha$ flux 
to estimate other magnitudes from this value. The electron density asummed 
is n$_{e}$ = 100 cm$^{-3}$ \citep{luri}.

The H$\alpha$ flux in a 15$\arcsec$ radius circular aperture (uncorrected for internal extinction) 
yields a lower limit to the luminosity $L_{H\alpha} \sim 2.3 \times 10^{40}$ erg s$^{-1}$.
The total number of H-ionizing photons, $Q(H^{0})$, estimated from \citet{oster} is:

\begin{displaymath}
Q(H^{0}) = \frac{\alpha_{B}}{\alpha^{eff}_{H\alpha}} \times \frac{L_{H\alpha}}{h\nu_{H\alpha}} 
\sim 2.2 \times 10^{52} s^{-1}
\end{displaymath}

\noindent where $\alpha_{B} / \alpha^{eff}_{H\alpha} \sim 2.99$, assuming case B recombination, 
$\alpha_{B} \sim 2.59 \times 10^{-13}$ cm$^{-3}$ s$^{-1}$, and an effective recombination coefficient 
for H$\alpha$ of 8.63 $\times$ 10$^{-14}$ cm$^{-3}$ s$^{-1}$ for 10$^{4}$ K 
and n$_{e}$ = 100 cm$^{-3}$ \citep{oster}. Thus the total (lower limit) ionized mass of the gas 
cloud is

\begin{displaymath}
M_{H^+} = Q(H^{0}) \frac{m_{p}}{n_{e}\alpha_{B}} \sim 7.3 \times 10^{5} M_{\odot}
\end{displaymath}

From evolutionary models of single stellar population ionizing clusters and radiation-bounded H\,{\sc ii} 
regions, a relation between $Q(H^{0})$ per solar mass and the H$\beta$ equivalent width can be found 
\citep{diaz}, which allows to estimate the stellar cluster mass by taking into account the cluster 
evolution:

\begin{equation}
log[Q(H^{0})/M_{\odot}]=0.86\,log[EW(H\beta)] + 44.48
\label{mion}
\end{equation}

Hence, a lower limit for the mass of the cluster can be estimated by means of the H$\beta$ equivalent 
width and the H$\alpha$ luminosity of the region. Using the range of values EW(H$\beta$)=134-296 \AA\ 
published by different authors \citep{luri, esteban, rosa94, rosa81}, and the value of Q(H$^{0}$) 
calculated above, equation \ref{mion} yields a mass of the ionizing stellar cluster in the range 
$0.6-1.1\times10^{6}$ M$_{\odot}$. Using the calibration of \cite{kenni98}
and the luminosity that we measure, $L_{H\alpha} \sim 2.3 \times 10^{40}$ erg s$^{-1}$, 
we obtain a star formation rate of 0.18 M$_{\odot}$ yr$^{-1}$, or 2.1 $\times$ 
10$^{-7}$ M$_{\odot}$ yr$^{-1}$ pc$^{-2}$ (taking the area of the region as the circular 
aperture defined in Figure \ref{apertura}), that translates into a cluster mass 
of $10^{6}$ M$_{\odot}$ for a burst duration of 10 Myr, which is the mean duration of the current 
burst implied by the analysis in the previous section. Nevertheless, it is known that this calibration 
was computed for a particular star formation episode at nearly constant rate during more than 100 Myr, 
so it might be not suitable for bursty systems. We have applied the new 
calibration by \cite{oti}\footnote{Their web tool can be found at 
\texttt{http://www.laeff.cab.inta-csic.es/research/sfr/}}, which makes a distinction 
between instantaneous (IB) and extended burst (EB). We used as input our measured H$\alpha$ 
luminosity, no color excess, and a Salpeter Initial Mass Function in the range 
1-100 M$_{\odot}$. For the case of instantaneous burst, \cite{oti} use the concept of 
star formation strength (SFS), defined as the total mass of gas transformed 
into stars during the burst, giving three different ages for the IB case. We obtained a SFS of 
1.5 $\times$ 10$^{6}$ M$_{\odot}$ for a burst of 4 Myr, 2.3 $\times$ 10$^{6}$ M$_{\odot}$ for 5 Myr, 
and 1.0 $\times$ 10$^{7}$ M$_{\odot}$ for the 6 Myr case. For a EB, the results are given in terms 
of the stardard star formation rate (SFR), with three different tabulated ages, namely: 
10, 30 and 250 Myr. 
However, in this age range the results are independent of the age for the H$\alpha$ estimator. 
If the region would be experiencing a continuous constant star formation, the SFR given by 
this calibration would be 0.1 M$_{\odot}$ yr$^{-1}$, less than a factor of 2 lower than the calibration 
by \cite{kenni98}\footnote{This calibration integrates a Salpeter IMF in the mass range 0.1-100 M$_{\odot}$, 
which predicts a larger mass.}. 

Thus we find that the masses of ionized gas and of the stellar cluster in the current star formation 
event are both of the same order of magnitud, close to $10^{6}$ M$_{\odot}$. This similarity may be 
understood in terms of a very high efficiency in star formation. Given that this is not the first 
event of star formation in NGC 5471 in the last 100 Myr, this high star formation efficiency takes an 
important meaning that implies that there has to be a large reservoir of gas
to maintain this important star formation activity. 

%--------------------------------------------- FIGURE FINAL CMD WITH ISOCHRONES AS COLOURED CONCAVE HULL
\begin{center}
\begin{figure*}
\includegraphics[width=1.\linewidth]{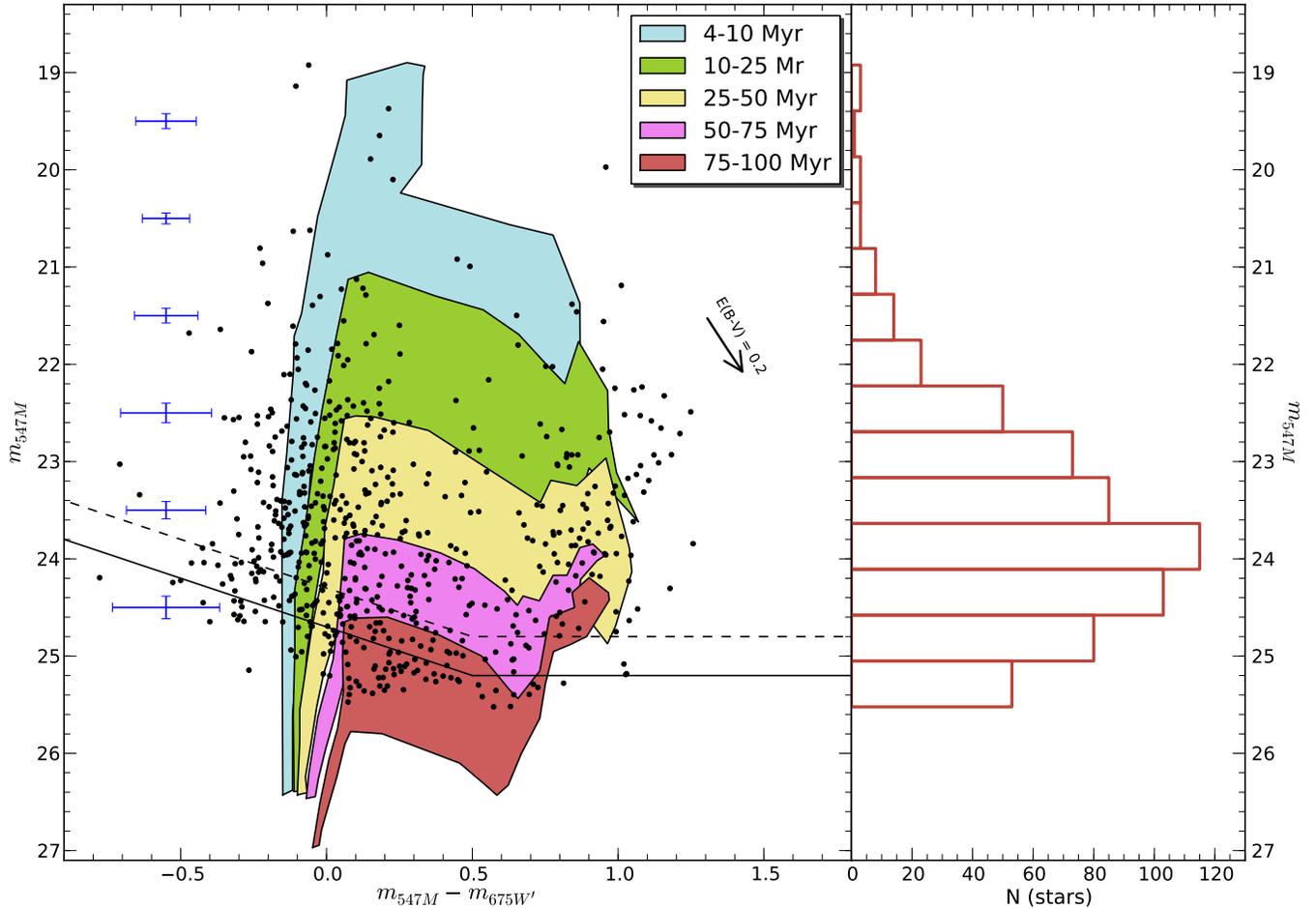}
\caption{
Final CMD with isochrones divided into five age bins, drawn by their concave hull, and shown with 
different colours. Error bars correspond to an average error in each magnitude bin. 
The right hand side of the figure shows the luminosity function.
The line across the bottom shows the completeness limit at the 50\% level in the core (dashed line) and 
in the rest of the image (solid line). 
}
\label{mpl_cmdiso_ms_lf}
\end{figure*}
\end{center}
%--------------------------------------------- FIGURE FINAL CMD WITH ISOCHRONES AS COLOURED CONCAVE HULL

Although optical images show NGC~5471 apparently isolated in the outskirts of M101, the H\,{\sc i} 
distribution of M101 shows that NGC 5471 is, in fact, one of the brightest regions in H\,{\sc i}, 
located in a spiral arm (cf. figure \ref{DSS_HI}). The thermal radio continuum emission from NGC 5471 
has the highest observed central emission measure of all the M101 complexes. Closely coincident with 
this H\,{\sc ii} radio continuum peak, there is an H\,{\sc i} concentration of nearly 
10$^{8}$ M$_{\odot}$ \citep{vialle}. The general distribution of the H\,{\sc i} gas in M101 
follows that of the diffuse FUV emission, with enhancements in the 21 cm line emission seen near 
FUV peaks. \citet{smith} include NGC~5471 as one of their 35 photodissociation regions (PDRs) 
candidates. Their Ultraviolet Imaging Telescope (UIT) FUV flux measurement is in good agreement with 
our GALEX measurement of $9.8 \times 10^{-14}$ erg cm$^{-2}$ s$^{-1}$ \AA$^{-1}$, 
and NGC~5471 is clearly the brightest FUV peak in the whole M101 galaxy. 
This is reflected in a decreased molecular content surrounding NGC~5471 which has been
dissociated by the high star formation efficiency, the highest in M101 as calculated by 
\citet{gian}.

%-----------------------------------------------------------------------------------------------------------------------------------------------------------------
\subsection{CMD}

The photometry of the individual stars presented in the color-magnitude diagram, 
m$_{547M}-$m$_{675W^{\prime}}$ (V-R) vs. m$_{547M}$ (V), of figure \ref{mpl_cmdiso_ms_lf}
is a powerful tool to further understand the star formation history in NGC 5471. 
The isochrones represented in the figure have been divided into five age bins,
(4-10, 10-25, 25-50, 50-75, 75-100) Myr, and drawn by their concave hull, shown with different colours.
The right hand side of the figure shows the luminosity function, which peaks at M$_{547M}=24$,
and becomes incomplete for fainter stars, as indicated by the 50\% completeness limit shown.
We recall that this CMD is the result of a careful process of calibration, measurement and cleaning,
as detailed in section \ref{resolved} above. We have not included in the CMD those point-like sources with a 
partially resolved PSF, that are likely to be compact clusters. At the distance of M101 some 
of the point sources are unresolved multiple systems, some of which may have been included in the CMD,
but these will not really affect any of the arguments that we present here.

A glance at the CMD tell us at once that star formation has been proceeding more or less continuously
for the last 100 Myr. The diagram is populated by stars of $\sim4-50$ M$_{\odot}$. 
We can readily identify clear structures in the diagram. 
A well-defined main-sequence at V-R $\sim$ -0.1, which begins with the brightest star at 19 mag 
(that corresponds to an age of $\sim4$ Myr) and goes all the way down to 25 mag. 
Red stars span a range of $3-4$ magnitudes in brightness, indicating that star formation has occurred 
during the last $\sim15-70$ Myr. 
There are some traces of the blue-plume at M$_V=-9$, indicating ongoing star formation. 
A strip of stars following a patch from [0.6,25] to [1.1,22.5] form the red giant branch. 

\subsubsection{Mass in stars and star formation history}

In order to quantify the relative importance of the present burst to the star formation history during 
the previous 100 Myr in NGC 5471, we make a rough estimate of the stellar mass from the CMD. We follow 
three different approaches. One method consists in assigning a mass to each of the stars in the CMD, 
corresponding to the most probable isochrone. 
This is done considering a 2D Gaussian probability density distribution centered on the 
distance in the color-magnitude space of a given observed star to the isochrones with 
dispersions being the estimated errors for the photometry. Once a mass has been assigned 
to a star, we count the stars in the mass range 6-55 M$_{\odot}$ and 
integrate a Salpeter IMF in the mass range 1-100 M$_{\odot}$. The result of this calculation gives 
$\sim2\times10^5$ M$_{\odot}$. The second method is as follows: first we assign a mean 
stellar mass of 6 M$_{\odot}$ to the stars in the most populated bin of the luminosity function, 
m$_{547M}=23.5$. This mass is then scaled up by the ratio of the total m$_{547M}$ of NGC 5471 to the 
magnitude of the bin m$_{547M}=23.5$. The resulting mass is used to normalize the IMF at the bin 
corresponding to 6 M$_{\odot}$. The result of integrating a Salpeter IMF in the range 1-100 M$_{\odot}$ 
thus normalized gives a value of $\sim5\times10^5$ M$_{\odot}$. 

%--------------------------------------------- FIGURE SFH BY STARFISH
\begin{center}
\begin{figure}
\includegraphics[width=1.\linewidth]{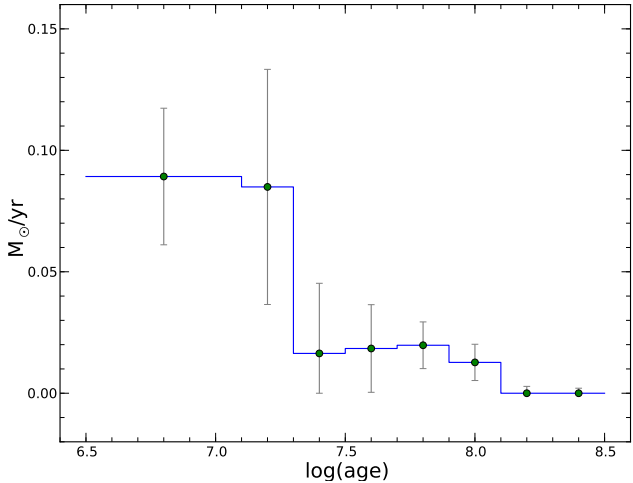}
\caption{
STARFISH recovered Star Formation History of NGC 5471.
}
\label{sfh-ngc5471}
\end{figure}
\end{center}
%--------------------------------------------- FIGURE SFH BY STARFISH

\begin{center}
\begin{figure}
\includegraphics[width=1.\linewidth]{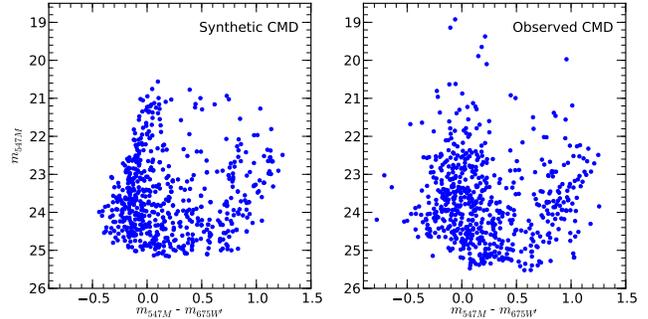}
\caption{
NGC 5471 synthetic CMD based on the best-fit SFH solution of STARFISH (\textit{left}) and the observed 
one (\textit{right}).
}
\label{synthcmd}
\end{figure}
\end{center}

A third independent method to estimate the mass is based on the fit of the CMD by means of Hess 
diagrams. We have used the code STARFISH\footnote{Available at 
\texttt{http://www.noao.edu/staff/jharris/SFH/}} 
\citep{harris01,harris02}, with the lastest patchsets, to find a best fit to the CMD 
by a linear combination of Hess diagrams at different ages. 
We have used our own set of isochrones described in section \ref{sbsec:iso}. The best-fit SFH of 
NGC 5471 is shown in Figure \ref{sfh-ngc5471}. Figure \ref{synthcmd} shows the synthetic CMD of NGC 5471 
constructed using the SFH amplitudes output of the best-fit SFH; the observed CMD is also shown for reference. 
The synthetic main sequence seems tighter and bluer compared to the observed one, probably due to 
differential reddening and underestimated errors.
The SFH derived seems to be consistent with a continuous one, but the errors only allow to take into 
consideration two episodes: a dominant young burst $<$ 20 Myr, preceded by a longer event during the 
period 20-100 Myr.
According to the SFR, the integrated mass in stars of the youngest event is comparable to the 
oldest one, with $\sim10^6$ M$_{\odot}$.

Although these methods give only an order of magnitude estimate, we can conclude that the integrated mass of
stars formed in the past 20-100 Myr is of the order of magnitude of the current star formation event,
that we have calculated above as $\sim10^6$ M$_{\odot}$. If the star formation proceeds in an isolated 
gas cloud, we would expect that the rate at which stars form would decrease with time, 
as less gas is left available to form new stars. This is what is observed in many giant H\,{\sc ii} 
regions with two or more generations of star formation, where the latest event is less massive than the 
previous. In the case of NGC 5471 the results from the integrated photometry and from the CMD indicate 
that the current star forming event is 
as massive as those during the previous 20-100 Myr. We can understand this in the context of the large 
H\,{\sc i} spiral arm in which the region is immersed (cf. section \ref{sec:massion}). In this 
respect, \cite{waller97} conclude from a morphological study of M101 that NGC 5471 may be the result of 
tidal interactions of M101 with the nearby galaxies NGC 5477 and NGC 5474 in the time interval 
100-1000 Myr ago.

%--------------------------------------------- FIGURE SFH BY STARFISH
%--------------------------------------------- FIGURE SFH BY STARFISH

%-----------------------------------------------------------------------------------------------------------------------------------------------------------------
\subsubsection{Spatial distribution of the star formation history}

%------------------------------------ FIGURE SPATIAL DISTRIBUTION OF CMD 4 BOXES
 \begin{figure*}[htbp]
 \begin{center}
$\begin{array}{cc}
\includegraphics[width=0.5\textwidth,angle=0]{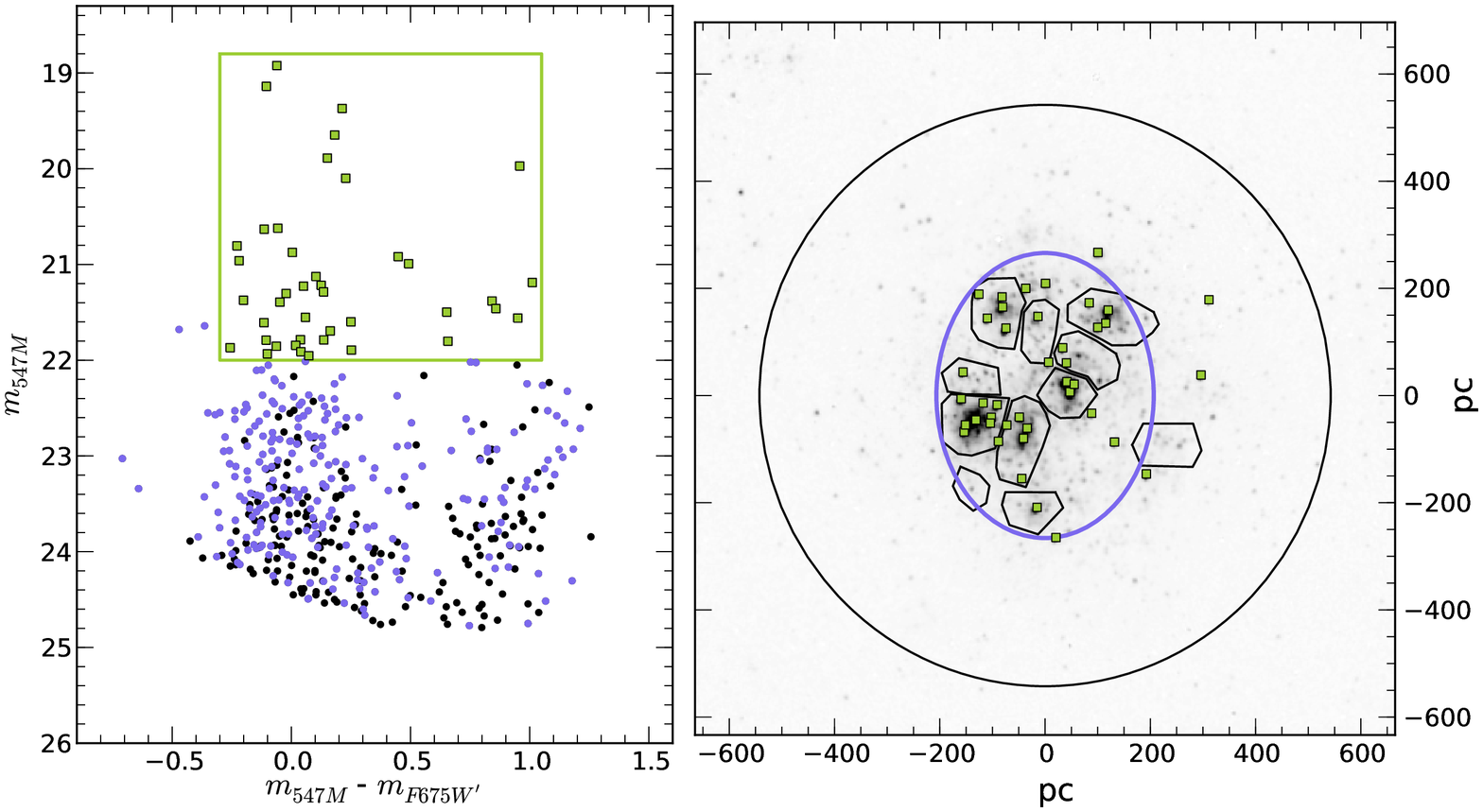} & 
 \includegraphics[width=0.5\textwidth,angle=0]{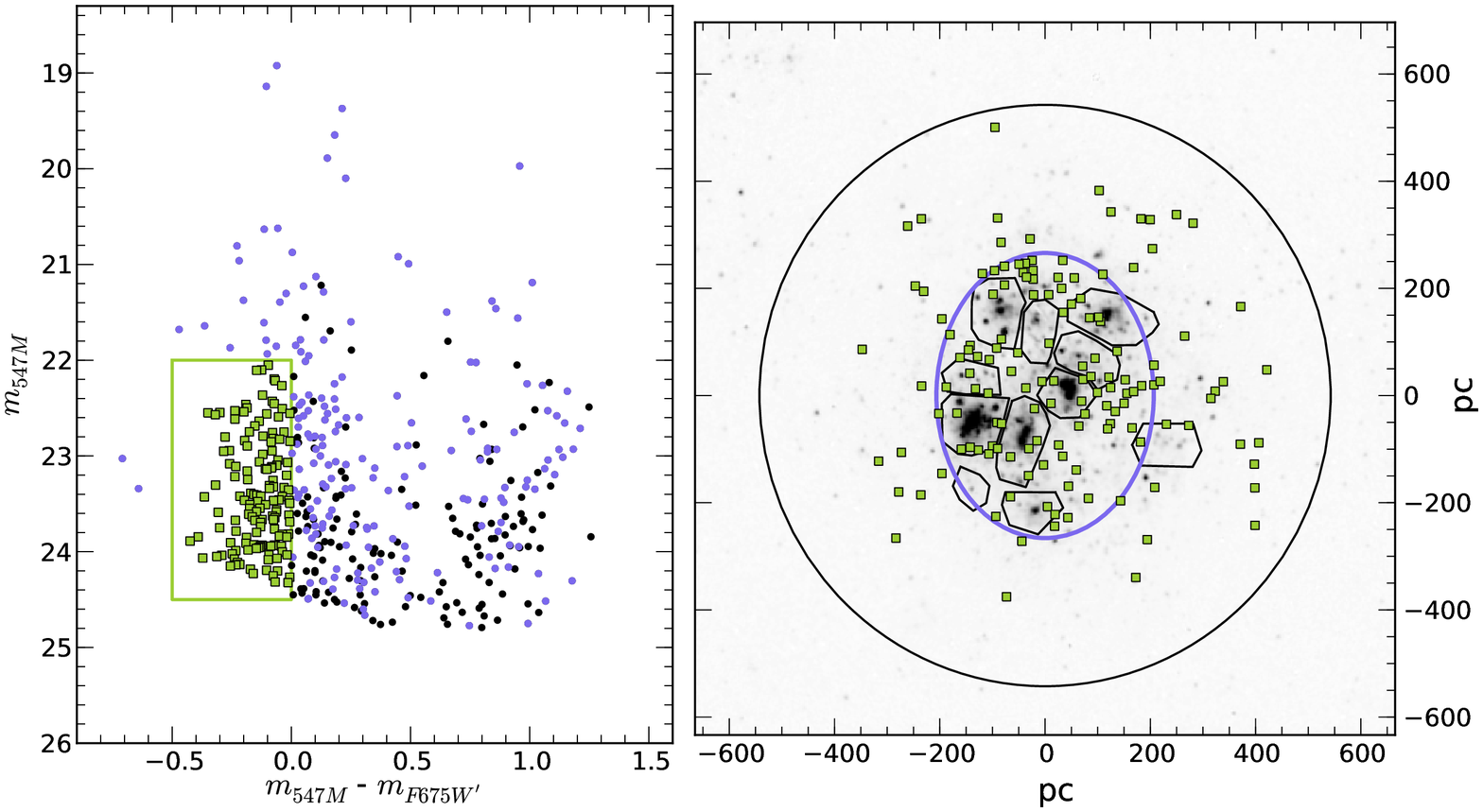} \\
\includegraphics[width=0.5\textwidth,angle=0]{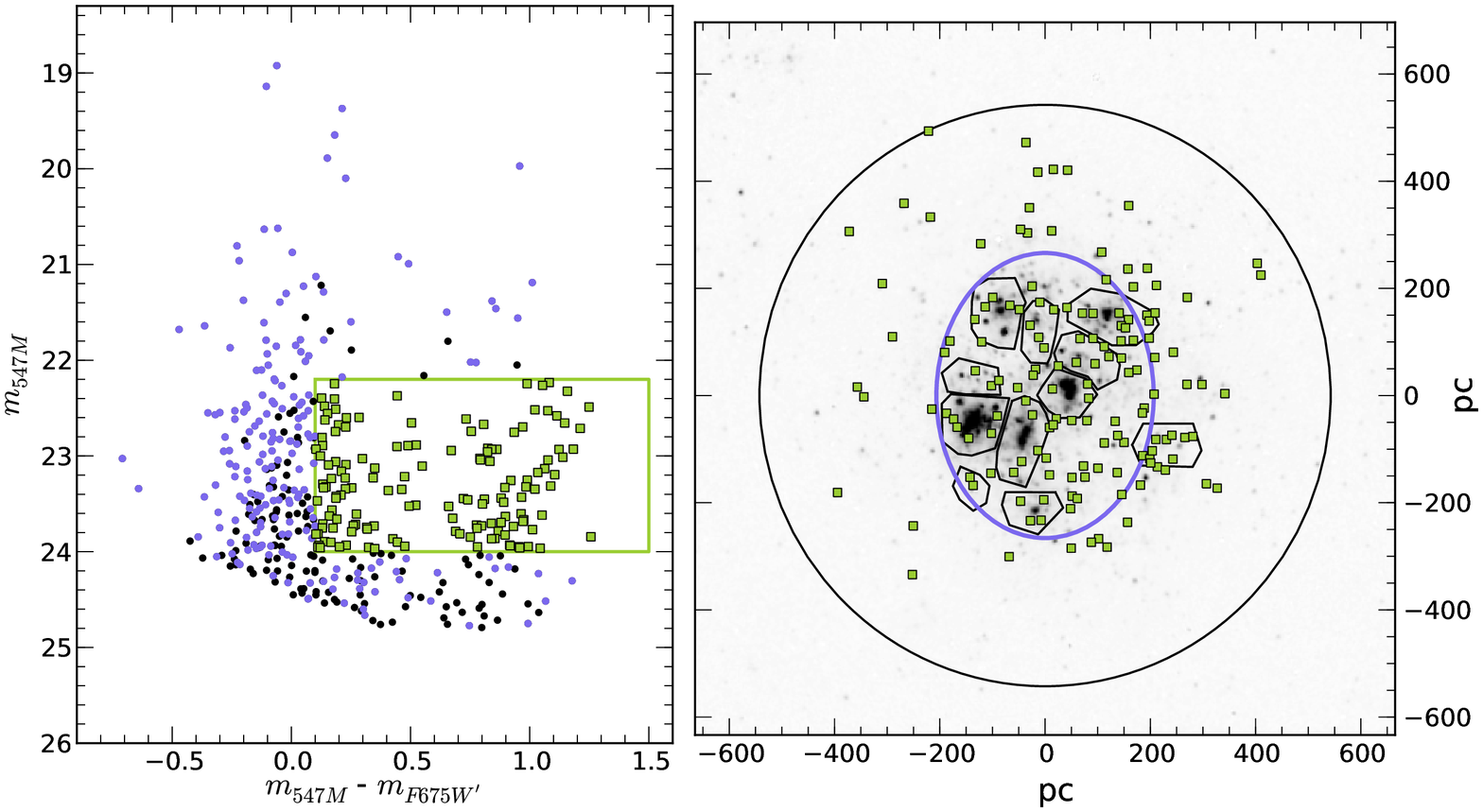} &
\includegraphics[width=0.5\textwidth,angle=0]{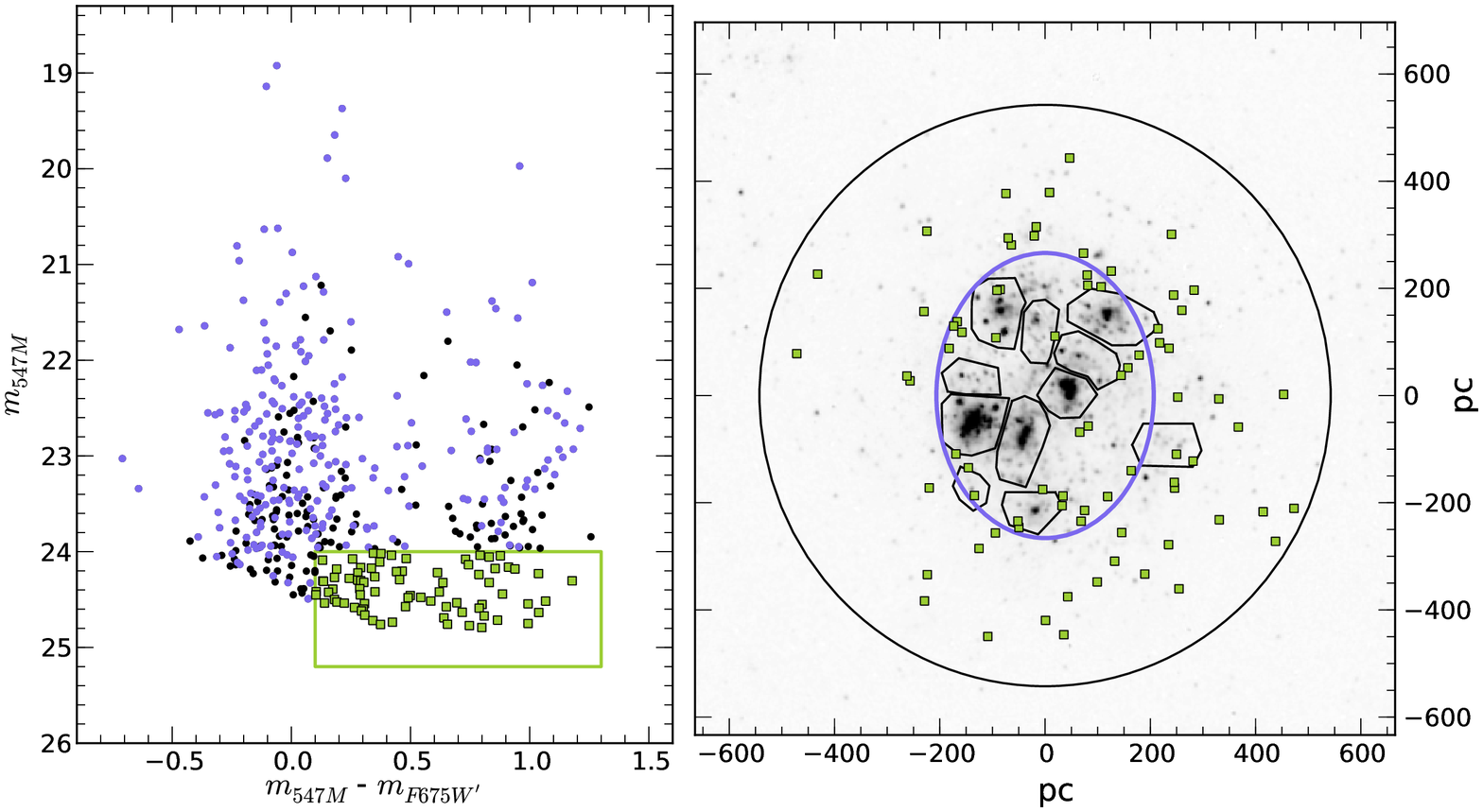} \\
\end{array}$    
\end{center}
\caption{Spatial distribution of the star formation with time. The CMD contains only stars with 
magnitudes below the 50\% completeness defined in the core region. The spatial distribution of the stars 
within four boxes selected in the CMD are shown in the right hand side images. The images are also 
drawn with a circle of 15 arcsec radius, indicating the area of the integrated photometry, an ellipse 
delineating the core region where the present star forming event occurs, and the polygons 
selected for the cluster analysis. The square points represent the stars in the box selected in the CMD, 
the light color points indicate the stars located in the core of the region (inner ellipse), while the 
black points stand for the stars belonging to the halo. Notice how the younger stars are concentrated in 
the inner clusters (top-left panel) and intermediate and older stars populate uniformily the region.} 
\label{fig:spatial_distrib_boxes}
\end{figure*}
%------------------------------------ FIGURE SPATIAL DISTRIBUTION OF CMD 4 BOXES

Given the long duration of the star formation in NGC 5471, $\sim 100$ Myr, and its large size $\sim 1$ kpc, 
it is natural to question the spatial location of the star formation through time along the extent of 
the region. In order to visualize this, we have selected four different areas in the CMD and proceeded 
to locate spatially the stars in these four boxes. Figure \ref{fig:spatial_distrib_boxes} 
displays the resulting distributions. 
The youngest and more massive stars (top-left panel) are clearly concentrated in the main star forming 
emission line knots, implying that massive stars are now forming mainly in clusters. Only two stars are not 
directly associated with these knots, one of which belongs to an ishocrone of 20 Myr.
The bottom-right panel displays the distribution of intermediate mass stars ($\sim5-10$ M$_{\odot}$) 
older than about 50 Myr; these stars are distributed mainly towards the halo, with only a few in the 
core of the region. 
The top-right and bottom-left panels show the distribution of intermediate mass stars 
($\sim10-15$ M$_{\odot}$), both those which are in the main sequence and those which have already 
evolved out of it. The distributions of stars in these two CMD boxes are clearly concentrated fairly 
uniformly throughout the core of NGC 5471; some of the stars in these two boxes are found in the halo, 
with a sparse distribution somewhat similar to the older lower mass stars in the bottom-right panel. 
Because the stars in these two intermediate boxes are of the same mass but they are found in a range 
of evolutionary stages, this implies that star formation has proceeded more or less uniformly in the 
core for the last $\sim20-50$ Myr.

\placefigure{spatial_distrib_boxes}

From this analysis of the spatial distribution of the CMD, a clear picture emerges in which the star 
formation in the complex NGC 5471 has proceeded in a general spatio-temporal sequence from the 
periphery to the core. During the first epoch, $\sim50-100$ Myr ago, the star formation occurred 
along the periphery and the core, then a second major event $\sim10-20$ Myr ago was more or less 
uniformly distributed mainly in the core, and finally the current ionizing star forming event is 
mostly concentrated in the singular bright clusters located inside the core, where the gas 
reservoir is presumably higher\footnote{Notice that the age dating results from the individual cluster 
photometry is compatible with this age-spatial distribution of the younger stars in the CMD, and this 
reinforces the age results obtained for the clusters.}. At the same time, we have seen from both 
the integrated photometry analysis and the CMD, that as the events proceeded concetrating from the 
periphery to the core clusters, the amount of mass of gas formed into stars seems to be of the same order 
of magnitude, but more concentrated in time, with higher star formation rate.

We note that the current event of ionizing clusters is apparently contained within a large bubble which 
defines the core of NGC 5471. This bubble, of projected size $400\times550$ pc, is clearly defined 
geometrically (cf. figure \ref{hst}) and kinematically \citep{cmt}, and is likely to have been produced 
by the stars that formed $\sim20$ Myr ago.\footnote{The results from a kinematical analysis of the expansion 
of this bubble are in progress and will be published elsewhere}

There are other examples that seem to follow the general process of star formation that we have outlined 
here for NGC 5471. \cite{ubeda2} find an inwards sequence of star formation and a bubble for cluster I-A 
in NGC 4214. \cite{wal97} find a triggered spatio-temporal distribution of stars in 30 Doradus, but this 
sequence occurs at the scale of the individual clusters in NGC 5471. Indeed there are indications that at 
this smaller scale, $\sim$50pc, the evolution of the different individual clusters indicated by the age 
dating is also reflected in their morphology (e.g. as mentioned in the case of 3+4), but a detailed study 
at these scales requires the high resolution UV-optical-IR imaging that can only be provided by the WF3 
camera onboard HST.

%-----------------------------------------------------------------------------------------------------------------------------------------------------------------
%-----------------------------------------------------------------------------------------------------------------------------------------------------------------
\section{Conclusions}
We have performed a photometric study of NGC 5471, a GEHR in the outskirts of the spiral galaxy M101, to 
derive its star formation history by means of resolved stellar photometry and integrated cluster photometry. 
Furthermore, integrated photometry integrated photometry of the whole region, using 
data from GALEX (ultraviolet), HST/WFPC2 (optical), and TNG (near infrared), yields two
possible solutions for a single stellar population: one that corresponds to an age of $\sim$8 Myr with
moderate extinction and one with an age around 60 Myr and very low extinction. This degeneration is typical
of this kind of approach when spatially integrated data are used and reflects the complexity of GEHR which
include clusters with different ages and extinction values.

From the photometric analysis of the eleven clusters defined on the IR H image a correlation emerges
in the sense that redder knots have less H$\alpha$ flux, showing a clear aging trend. The ages range 
from 3 Myr for the youngest cluster, up to 10 Myr for the oldest one. However, the Lowest Luminosity 
Limit test prevents us from taking these estimates at face value, since only a few individual clusters 
satisfy the required limit.

The complex history of star formation of NGC~5471 revealed by the cluster analysis is confirmed by the
resolved stellar analysis. From the CMD it is clear that star formation has been proceeding 
continuously with more or less intensity for the 
last 100 Myr. The well-defined main-sequence gives a youngest age of about 4 Myr,
while red stars, which span a range of 3-4 magnitudes, indicate that star formation has occurred during
the last 15-70 Myr.

We have found that the masses of the ionized gas and of the stellar cluster in the current star 
formation event are both of the same order of magnitude, around 10$^{6}$ M$_{\odot}$, which may 
be understood in terms of a very high efficiency.
In fact, the star formation strength of NGC 5471 is more than a factor of 2 that of 
NGC 604 in M33 \citep{relano}. This implies that there has to be a large reservoir of gas to 
maintain this star formation rate, something that is confirmed by the huge H{\sc i} concentration 
in which the region is located.  

From the spatial distribution of the stars, we can conclude that the star formation in NGC~5471 has
proceeded in a general spatio-temporal sequence which was more extended in the past and now is 
concentrated in clusters in the core. The current event of ionizing clusters is contained within
a large bubble, which is likely to have been produced by the stars that formed $\sim$ 20 Myr ago. 
This leads to the general question of wether a low density ``seed'' generation of evolved intermediate 
mass stars is in fact a prerequisite to trigger a starburst event \citep{brandner02}. 
More UV-optical-IR data at higher spatial resolution, such as that can be provided by the WF3 camera on 
board HST, are obviously needed in order to have a conclusive answer. 

%-------------------------------------------------------------------------------

\acknowledgments

We have benefited from comments and suggestions by Emilio Alfaro, David Mart\'inez Delgado, 
and Nolan Walborn. This work has been funded by DGICYT programs AYA2004-02703, AYA2007-64712, 
AYA2004-08260-C03-03, and  AYA2007-67965-C03-03. RGB acknowledges support from the Spanish 
MEC through FPI grant BES-2005-6910 and from the China National Postdoc Fund 
Grant No. 20100480144. This work has made use of data from the archives of
NASA/ESA Hubble Space Telescope, obtained at the Space Telescope Science
Institute, which is operated by the Association of Universities for
Research in Astronomy, Inc., under NASA contract NAS5-26555, and of data
from the Telescopio Nazionale Galileo, operated on the island of La Palma
by the Galileo Galilei Foundation INAF in the Spanish Observatorio del
Roque de Los Muchachos of the Instituto de Astrof\'\i sica de Canarias.
Thanks to cheerful Francesca Ghinassi for her help during the TNG observations.
We thank an anonymous referee for very useful comments that improved the presentation 
of the paper.

%--------------------------------------------------------------------------------------------------------------------------------------------------
%--------------------------------------------------------------------------------------------------------------------------------------------------

\end{document}